\documentclass{aa}  
\usepackage{amsmath}
\usepackage{comment}
\usepackage{lineno}
\renewcommand\linenumbers{}
\let\linenumbers\nonumber

\AtBeginDocument{\nolinenumbers}
\usepackage{graphicx}
\usepackage{txfonts}
\usepackage{color}
\usepackage{xcolor}
\usepackage[colorlinks]{hyperref}

\hypersetup{
    colorlinks = true,
    linkcolor = {blue},
    citecolor = {blue},
}

\usepackage{array}
\usepackage{multirow}
\newcolumntype{P}[1]{>{\centering\arraybackslash}p{#1}}

\begin{document}

   \title{A Novel Method to Estimate the FUV Flux and a Catalogue for Star-Hosting Discs in Nearby Star-Forming Regions}


   \author{Rossella Anania
          \inst{1}, 
          Andrew J. Winter \inst{2},
          Giovanni Rosotti
          \inst{1},
          Miguel Vioque
          \inst{3},
          Eleonora Zari,
          \inst{4},
          Michelangelo Pantaleoni González, 
          \inst{5}
          Leonardo Testi
          \inst{6}
          }

   \institute{Dipartimento di Fisica, Università degli Studi di Milano, Via Celoria 16, I-20133 Milano, Italy\\
              \email{rossella.anania@unimi.it}
        \and
            Université C\^ote d'Azur, Observatoire de la C\^ote d'Azur, CNRS, Laboratoire Lagrange, 06300, Nice, France
        \and
            European Southern Observatory, Karl-Schwarzschild-Str. 2, 85748 Garching bei München, Germany
        \and 
            Dipartimento di Fisica e Astronomia, Università di Firenze, Via G.
Sansone 1, I-50019, Sesto F.no (Firenze), Italy
        \and
            Department of Astrophysics, University of Vienna, Türkenschanzstrasse 17, 1180, Vienna, Austria
        \and 
            Dipartimento di Fisica e Astronomia, Università di Bologna, Via Gobetti 93/2, 40122, Bologna, Italy; INAF – Osservatorio Astrofisico di Arcetri, Largo E. Fermi 5, 50125, Firenze, Italy
             }

   \date{Received XXX; accepted YYY}

 
  \abstract{Protoplanetary discs, when externally irradiated by Far Ultraviolet (FUV) photons from OBA-type stars, lose material through photoevaporative winds, reducing the amount of material available to form planets. Understanding the link between environmental irradiation and observed disc properties requires accurately evaluating the FUV flux at star-hosting discs, which can be challenging due stellar parallax uncertainties. In this paper, we address this issue proposing a novel approach: using the local density distribution of a star-forming region (i.e. 2D pairwise star separations distribution) and assuming isotropy, we infer 3D separations between star-hosting discs and massive stars. We test this approach on synthetic clusters, showing that it significantly improves the accuracy compared to previous methods. 
  We compute FUV fluxes for a large sample of star-bearing discs in 7 regions within $\sim$200 pc, 6 regions in Orion, and Serpens sub-regions,  providing a publicly accessible catalogue. We find that discs in regions hosting late-type B and early-type A stars can reach non-negligible irradiation levels for disc evolution (10-100 $\mathrm{G}_{0}$).
  We investigate dust disc masses relative to FUV fluxes detecting hints of a negative correlation when restricting to average region ages. However, we emphasize the need for more stellar and disc measurements at >$10^{2} \ \mathrm{G}_{0}$ to probe the dependence of disc properties on the environmental irradiation. Including average interstellar dust extinction, median FUV fluxes are not significantly attenuated, though this result may change if high-resolution 3D dust extinction maps were available. The method presented in this work is a powerful tool that can be expanded to additional regions.
  }
   \keywords{protoplanetary discs – accretion, accretion discs - stars: distances }

   \titlerunning{ }
   \authorrunning{Anania, R. et al.}
    
   \maketitle

%

\section{Introduction}
The majority of young stars (and their surrounding protoplanetary disc) form in dense and massive clusters or associations (e.g. \citeauthor{miller_scalo_1978} \citeyear{miller_scalo_1978}, \citeauthor{Lada_2010} \citeyear{Lada_2010}), where the Ultraviolet (UV) radiation emitted by massive stars, primarily of spectral type O and B, can strongly influence disc evolution and planet formation, as well as their physical and chemical composition. 
In \citet{Vioque_2023} is shown that massive star-forming regions become less clustered over time, with no preferential position occupied by massive stars in the cluster.  Therefore, as protoplanetary discs initially distant from massive stars may still experience intense irradiation over time, it is essential to be aware of the distribution and characteristics of massive stars in star-forming regions where these discs are observed. 

The intense Extreme Ultraviolet (EUV) and Far Ultraviolet (FUV) radiation from OB stars heats the outermost disc regions, weakly gravitationally bound to the central star, leading to an outside-in depletion of material.
This process, known as external photoevaporation, can result in smaller, less massive, and shorter-lived discs (see \citeauthor{winter_photoev} \citeyear{winter_photoev} for a review, and references therein). 
The dust component, depending on its size and composition, can also be affected by photoevaporation, with dust grains being trapped in thermal winds and carried away from the disc (e.g. \citeauthor{facchini_2016} \citeyear{facchini_2016}, \citeauthor{Sellek_dust_mass_2020} \citeyear{Sellek_dust_mass_2020}).
Disc truncation due to photoevaporation prevents the disc from spreading viscously, and the loss of gas and dust reduces the location and material available for forming new planets (\citeauthor{sellek_fried} \citeyear{sellek_fried}, \citeauthor{Quiao_coleman_2023} \citeyear{Quiao_coleman_2023}). As grain growth and formation of gas giants in the outermost disc regions can be limited by photoevaporation, this will result in a different observed exoplanet population (e.g. \citeauthor{Ndugu_2018} \citeyear{Ndugu_2018}).

Evidence of external photoevaporation in action on protoplanetary discs is probed by direct observations of proplyds, which are protoplanetary discs seen in optical lines as ionised gas clouds with a characteristic comet-like shape, with the cusp pointing toward the nearest massive star (e.g. \citeauthor{ODell_1993} \citeyear{ODell_1993}, \citeauthor{Otter_2021} \citeyear{Otter_2021}, \citeauthor{Berne_2024} \citeyear{Berne_2024}, \citeauthor{Aru_2024} \citeyear{Aru_2024}). These fascinating objects are found in highly irradiated environments ($\gtrsim 3 \times 10^3 \ \mathrm{G}_{0}$\footnote{The FUV flux is given in terms of the Habing unit, G$_{0}$ \citep{Habing_G0}, which quantifies the flux integral over the range of wavelengths [912 - 2400] $\rm{\mathring{A}}$, normalised to the average flux in the solar neighbourhood ($1.6\times 10^{-3}$ erg s$^{-1}$ cm$^{-2}$)}, \citeauthor{Kim_proplyd_2016} \citeyear{Kim_proplyd_2016}), in close proximity of massive O-type or early-type B stars (within $\sim$0.1 pc).
However, proplyds are representative only of the most extreme environmental conditions, while the majority of discs form and evolve in moderately irradiated environments ($10^{2} - 10^{3} \ \mathrm{G}_{0}$, \citeauthor{Fatuzzo_Adams_2008} \citeyear{Fatuzzo_Adams_2008}), where external irradiation is driven by less massive stars, and the peculiar proplyd-like shape is not detectable.
In intermediate irradiated environments, the effect of photoevaporation is indirectly suggested by the dependence of the (sub-)mm continuum flux emission with the distance from the most massive star, such as in $\sigma$ Orionis (\citeauthor{Ansdell_2017_sori} \citeyear{Ansdell_2017_sori}, \citeauthor{Mauco_2023_sori} \citeyear{Mauco_2023_sori}), and $\lambda$ Orionis \citep{Ansdell_lambdaOri}. The low-irradiated extended population of Orion investigated by the SODA Survey ($\sim$ 1-100 $\mathrm{G}_{0}$) also show a correlation between (sub-)mm continuum flux emission and FUV flux \citep{SODA_2023}. Moreover, a weak correlation between gas and dust disc sizes and FUV flux is even found for the AGE-PRO sample in the Upper Scorpius region (2-12 $\mathrm{G}_{0}$, \citeauthor{AGEPRO_VIII_ext_photoevap}\citeyear{AGEPRO_VIII_ext_photoevap}), where adding external photoevaporation to a viscous model results in better reproducing gas disc sizes.

Depending on strength and intensity of the incident UV photons, the thermal and chemical structure of protoplanetary discs can also be altered, particularly in the most external surface layers \citep{Walsh_2013}, while recent results from the eXtreme UV Environments (XUE) James Webb Space Telescope
(JWST) program show that the chemistry of the inner disc (< 10 au) is similar between lower and higher irradiated discs \citep{Macla_JWST}. 
Astrochemical models predict that the outer disc molecular composition should be influenced by external irradiation (e.g. \citeauthor{Nguyen_2002} \citeyear{Nguyen_2002}, \citeauthor{Walsh_2013} \citeyear{Walsh_2013}, \citeyear{Walsh_2014}). The impact of intense UV fields on the molecular composition of the outermost disc regions is supported by observational evidence on highly UV-irradiated discs (e.g. CH$_{3}^{+}$ detection in a $\sim 10^{5} \ \mathrm{G}_{0}$ irradiated disc \citeauthor{Berne_ch3_photoev}\citeyear{Berne_ch3_photoev}).
However, at moderate and typical UV levels (<$10^{3} \ \mathrm{G}_{0}$),  direct detection is currently unavailable.

All the above results highlight that the environment plays a crucial role in shaping disc physical and chemical properties, even at moderate levels of irradiation. Therefore, accurately quantifying the FUV flux at the position of star-hosting discs is essential to reliably compare the predictions of disc evolution models and astrochemical models with observational outcomes, and investigate the role of the environmental FUV irradiation on disc and planetary evolution. 
However, FUV flux can be extremely challenging evaluating as it requires knowledge of the distribution of stars in a three-dimensional (3D) space, which is highly influenced by parallax uncertainties. The separation value typically used in flux calculations is the 2D separation projected on the sky plane, assuming an average distance to the entire cluster. This precludes an estimate of uncertainty on the distance between stars, and therefore on the FUV flux to which a given protoplanetary disc is exposed.

In the work of \citet{FUV_from_dust}, they provided an FUV flux map of the Orion region by estimating stellar luminosity under the assumption that it is entirely re-radiated in the dust continuum, observed in infrared wavelengths.
The FUV flux was calculated applying a $\sim$constant ratio between FUV luminosity and stellar luminosity for early OB-type stars, and, also in this case, using 2D projected separation distances from the most massive stars. 
Moreover, late-type B and early-type A stars, which are usually neglected in the FUV flux calculation, can significantly influence the evolution of protoplanetary discs if located 
close to them.
\\

In this work we aim to provide a more precise estimate of the FUV flux, along with its uncertainty, at the position of star-hosting discs in nearby star-forming regions. Specifically, we propose using the available information on the 2D geometry of a star-forming region to constrain the 3D separation between massive stars and star-hosting discs, which is the largest source of uncertainty in the flux calculation. 
We provide a publicly available catalogue of the FUV flux experienced by star-hosting discs located in nearby regions presenting previous studies of disc populations (see Table \ref{tab:big_table_of_fluxes}), covering regions within $\sim$200 pc that have been well-investigated by protoplanetary disc surveys (i.e. Taurus and Lupus), well-known Orion star-forming regions (i.e. ONC and $\lambda$Orionis), and Serpens sub-regions at average distance of $\sim 500$ pc. 
The selected disc sample and the star-forming regions included in this work are discussed in Sec. \ref{sec:sample_selection}. 
In Sec. \ref{sec:fuv_calculation:sample_star} we describe the selection criteria for the irradiating massive stars. In Sec. \ref{sec:fuv_calculation_methods_dist_eval} we introduce three approaches to evaluate the FUV flux accounting for the uncertainty in line-of-sight distance, and we test the methods on synthetic clusters in Sec. \ref{sec:synthetic_clusters}.
The results for the various star-forming regions are presented in Sec. \ref{sec:results}, along with an investigation on the relation between external UV radiation and measured dust disc masses, in Sec. \ref{subsec:dust_masses}. In Sec.\ref{sec:discussion} we discuss the results obtained and the uncertainties and future improvement of the method. 
We give our conclusions in Sec. \ref{sec:conclusions}. 
The online Table with the FUV flux estimate at the position of numerous discs in the regions studied, is available in Appendix \ref{appendix:big_table}, Table \ref{tab:big_table_of_fluxes}. 

\section{Target regions and 
discs}\label{sec:sample_selection}

We evaluated the FUV flux at the position of star-hosting discs located in nearby star-forming regions ranging distances from Taurus ($\sim$140 pc) to Serpens ($\sim$500 pc). 
\begin{table}[h]
\caption{\label{tab:region_ref} Star-forming regions considered in this study and corresponding disc samples.}
    \centering
    \vspace{0.3cm}

    \begin{tabular}[H]{|P{2.4cm} | P{1.1cm} | P{2.2cm} | P{1.1cm} |}

    \hline
    \hline

    Region & median distance [pc] & reference disc sample & number of discs included \\

    \hline
    
    Upper Sco & 142 & \citet{Manara_2023} & 113 \\
    Lupus & 160 & \citet{Manara_2023} & 100 \\
    $\rho$ Oph & 140 & \citet{Manara_2023} & 279 \\
     Taurus & 140  & \citet{Manara_2023} & 217 \\
    Chamaeleon I & 179 & \citet{Manara_2023} & 93\\
    Chamaeleon II & 181 & \citet{Manara_2023} & 41\\
    Corona Australis & 150 & \citet{Manara_2023} & 48\\
    ONC & 414 & \citet{Ricci_2008_proplyds}, \citet{Eisner_2018_onc} & 368 \\
    $\sigma$ Orionis & 385 & \citet{Ansdell_2017_sori}, \citet{Mauco_2023_sori} & 50 \\
    $\lambda$ Orionis & 450 & \citet{Ansdell_lambdaOri} & 44 \\
    NGC 2024 & 414 & \citet{VanTerwisga_ngc2024} & 179 \\
    NGC 1977 & 414 & Kim et al. in prep. & 79 \\
    25 Orionis & 356 & \citet{Briceno_19_25ori} & 64 \\
    Serpens & 350-500 & \citet{Anderson22_Serpens} & 245 \\
    \hline
    
    \end{tabular}
    \tablefoot{ For each region included in this work, we indicate median distance, references for the disc sample considered, and total number of discs in the disc sample.
    }
    
\end{table}
In order to explore the level of FUV irradiation experienced by discs in various environmental conditions, we included regions hosting a different number of massive stars covering a wide range of spectral types, and featuring well-known populations of disc-bearing stars with estimated age $\lesssim$10 Myr. The investigated regions are thus motivated by previous studies of disc populations.
The rationale behind this selection is to present and apply our method for calculating the FUV flux and its associated uncertainty.
By doing so, we access both the irradiation on individual discs, and a statistical investigation across different regions. 
In particular, we selected discs from seven of the most nearby star-forming regions, which have been intensively investigated by disc surveys (\citeauthor{Manara_2023} \citeyear{Manara_2023} and references therein).
These are Upper Scorpius (Upper Sco), Lupus, Taurus, $\rho$ Ophiuchi ($\rho$ Oph), Chamaeleon I (ChamI), Chamaeleon II (ChamII), and Corona Australis (CrA). 
In addition, we included discs from six star-forming regions in Orion, which are the Orion Nebula Cluster (ONC), $\sigma$ Orionis ($\sigma$ Ori), $\lambda$ Orionis ($\lambda$ Ori), 25 Orionis (25 Ori), NGC 2024, and NGC 1977, where more luminous O-type and early B-type stars are located. Furthermore, the Serpens region is divided in four clusters of disc based on \citet{Anderson22_Serpens}.

The number of discs considered in each region, and the references for the selected disc sample are listed in Table \ref{tab:region_ref}. In particular, we refer to \citet{Manara_2023} for the disc sample in the most nearby regions ($\lesssim$ 200 pc), and to \citet{Anderson22_Serpens} for the ClassII Serpens discs. The discs in the Orion region are taken from various catalogues (see Table \ref{tab:region_ref}). Specifically, the disc sample in the 25 Ori region consists of the sources in \citet{Briceno_19_25ori} presenting counterpart in the AllWISE catalogue \citep{AllWISE_catalogue} and infrared excess according to 
\citet{Luhman_scocen_discs} being consistent with the classification of full, transitional, or evolved discs \citep{Esplin_2018}. 
For the ONC, we selected the discs located in the innermost region around the Trapezium cluster, as included in the work by \citet{Eisner_2018_onc}, and completed this initial catalogue with the proplyds listed in the catalogue by \citet{Ricci_2008_proplyds}. Future investigations will expand this initial catalogue to include discs located in the outermost part of the region.

\section{FUV Flux calculation: massive hot stars and their properties}\label{sec:fuv_calculation:sample_star}
The total FUV flux in a certain position is usually computed adding the contribution of individual neighbouring massive stars:
\begin{equation}
    F_{\mathrm{FUV, disc}} = \sum_{m} \frac{L_{\mathrm{FUV,m}}}{4 \pi |x_{\mathrm{disc}} - x_{\mathrm{m}}|^{2}},
    \label{eq:fuv_flux}
\end{equation}
where $L_{\mathrm{FUV,m}}$ is the FUV luminosity of the massive star and $|x_{\mathrm{disc}} - x_{\mathrm{m}}|$ is the separation between the location at which we want to evaluate the flux (in our case the disc position) and the massive star. 
In the calculation of the FUV flux the selected massive stars are often limited to the nearest (or a few nearest) massive stars in the field, assuming that the closest stars dominate the flux, while the influence of more distant stars is typically neglected. However, including high and intermediate-mass stars that are not members of a specific star-forming region increases the completeness of the analysis. In our study, we used a large sample of massive stars, enabling us to explore the impact of irradiation on discs in regions with fewer massive stars.
The criteria employed for selecting the massive stars that contribute to the overall FUV flux on individual discs are discussed in Sec. \ref{subsec:OB_selection}, while Sec. \ref{subsec:fuv_luminosity} describes how we retrieved the FUV luminosity.

\subsection{Massive Stars Selection} 
\label{subsec:OB_selection}

In order to perform a nearly comprehensive selection of massive stars that influence the total FUV flux on discs in specific regions, we cross-matched multiple stellar catalogues, as follows:
\begin{itemize}
    \item ALS III catalogue: we selected from the ALS III catalogue (Pantaleoni-Gonzalez et al., in prep.) all the massive stars in 2 kpc from the Sun having good Gaia DR3 \citep{Gaia_dr3}\footnote{\url{https://www.cosmos.esa.int/web/gaia/dr3}} photometric and spectrometric measurements\footnote{Stars which present nebular contamination in the Gaia colour-colour diagrams, miss data in one of the 3 Gaia bands, have RUWE > 1.4 or missing parallax values are excluded from this sample}. This first selection provides a pure list of the known, confirmed, most massive stars having accurate Gaia measurements and includes 5791 stars covering spectral type O, B2 and earlier for Main-Sequence stars, B5 and earlier for Giants, B9 and earlier for the Super-Giants, Wolf Rayet (WR) stars and stars lacking spectral type but classified as massive since they are situated above the 20 kK extinction track in the HR diagrams. \\
    
    \item \citet{Zari_catalogue} catalogue: we cross-matched the initial ALSIII sample with the massive stars included in the catalogue of hot stars of \citet{Zari_catalogue}, closer than 2 kpc from the Sun, and we added the stars with effective temperature greater than $\sim$8000 K (corresponding approximately to a A5V star) non included in our first sample. In total, we included 218648 hot-stars from this catalogue. \\
    
    \item Gaia ESP-HS catalogue: we added the stars (if not already included in the previous catalogues) contained in the Gaia DR3 Extended Stellar Parametrizer for Hot Stars pipeline (ESP-HS, \citeauthor{GaiaDR3_ESPHS} \citeyear{GaiaDR3_ESPHS}, \citeauthor{GaiaDR3_ESPHS_2} \citeyear{GaiaDR3_ESPHS_2}), which includes stars classified as A, B and O based on the Extended Stellar Parametrizer for Emission-Line Stars classifier (ESP-ELS). This sample also includes the stars that we excluded from the ALSIII initial selection due to their uncertainty or unknown parallax measurement, but that are classified as OBA-type.    
    From this last sample we exclude the outliers, which are stars with effective temperature associated to the parameter \texttt{teff\_esphs} higher than 23000 K, presenting large error bars and not reliable measurements. \\
    
    \item Hipparcos catalogue: as some of the most luminous stars are not included in the Gaia DR3 catalogue, due to the challenging astrometric and photometric measurements, we included the stars of type O, B, A0, and A1 contained in the Hipparcos catalogue (\citeauthor{Hipparcos_catalogue} \citeyear{Hipparcos_catalogue}, \citeauthor{Hipparcos_2007} \citeyear{Hipparcos_2007}) \footnote{\url{https://www.cosmos.esa.int/web/hipparcos/home}} that are not included in the previous selection. The sample of massive stars selected from Hipparcos consists of $\sim$6000 stars.
    
    \item SIMBAD database: for each star-forming region, we checked and included the stars lacking temperature estimate in Gaia, but classified as O, B, A0 or A1 in SIMBAD with a well-defined luminosity class and reference.
\end{itemize}

Including in the FUV flux calculation not only the most luminous and massive stars, but also late B-type and early A-type stars is particularly important in regions where no O-type and early B-type stars are found but the environmental irradiation field is not negligible (e.g. Upper Sco). Indeed, while the average FUV flux is dominated by the most massive and luminous stars, late B-type and early A-type stars in the close proximity of protoplanetary discs could also be relevant in determining the FUV flux they experience and, consequently, in shaping disc properties.




\subsection{Calculation of the FUV Luminosity}\label{subsec:fuv_luminosity}
The FUV luminosity of the massive stars, which is a crucial parameter for evaluating the FUV flux, is evaluated employing the following procedure.
We derived the effective temperature from the spectral type and luminosity class of the massive stars using the spectral classification given by the Galactic O-Star Catalogue (GOSC, \citeauthor{GOSC_catalogue} \citeyear{GOSC_catalogue}), when available. This classification is based on high resolution spectra other than only on photometric measurements.
For the stars lacking GOSC spectral classification and contained in the ESP-HS Gaia DR3 pipeline, we used the effective temperature addressed by the parameter \texttt{teff\_esphs} and the associated uncertainty. This is a more reliable estimate of the effective temperature for OBA stars, which is performed by the ESP-HS package, and is characterised by the omission of the regions of the BP/RP and RVS spectra dominated by emission lines that influence the calculation of the temperature.
Alternatively, we use the parameter \texttt{teff\_gspphot}, which corresponds to a temperature based on the assumption that the entire spectrum has a temperature-based origin in the stellar photosphere. The temperature of stars lacking this two parameters, and the Hipparcos stars, is retrieved from their spectral classification
through tables referring to Main-Sequence stars, Giant, Super-giant and WR stars (\citeauthor{Pecaut_Mamajek_2013} \citeyear{Pecaut_Mamajek_2013}, \citeauthor{Gray_Corbally_2009} \citeyear{Gray_Corbally_2009}). 
After evaluating the effective temperature, we used the isochrones provided by the MIST models (\citeauthor{Mist_1} \citeyear{Mist_1}, \citeauthor{Mist_2} \citeyear{Mist_2}), assuming a fixed age (we used 1 Myr), to obtain the stellar parameters (stellar mass $M_{\star}$, radius $R_{\star}$, luminosity $L_{\star}$ and the surface gravity $g$) needed to extract the stellar flux from the \citet{Castelli_Kurucz_2004} ATLAS9 stellar atmosphere models. Finally, the stellar flux is integrated in the FUV range of wavelengths [912 - 2400] $\mathring{\rm{A}}$ to retrieve the FUV luminosity.

The method described above requires assuming an age for the massive stars, as the stellar FUV luminosity increases during a star's lifetime. However, the work of \citet{Kunitomo_2021} demonstrated that intermediate and high-mass stars present a constant FUV luminosity after $\sim$1 Myr of evolution. Therefore, setting the stellar age to 1 Myr does not significantly influence the final result.

\section{FUV flux calculation: accounting for the uncertainty on the 3D separation }\label{sec:fuv_calculation_methods_dist_eval}

The FUV flux calculation is significantly influenced by the uncertainty on the true position of the stellar objects in a three-dimensional space.
Indeed, even a small error on the relative separation between discs and massive stars can result in large uncertainties on the final flux values, due to the strong dependence of FUV flux on the relative distance. This is particularly relevant the more discs are located closer to luminous stars, where errors on the relative distance between the two objects (caused by uncertainties in parallax measurements) can be comparable or exceed their separation.
In this work, we addressed this issue by presenting and comparing three approaches to evaluate the FUV flux, taking into account the uncertainty in the line-of-sight distance of the stars from the observer (measured from parallax), which is known with far less precision compared to the RA and Dec components:
\\
\\
Method 1: Distance uncertainty sampling. The radial distance of each stellar object is evaluated performing a Monte Carlo sampling of the distribution of distances based on the Gaia DR3 and Hipparcos parallax measurements and uncertainties, as described in Sec. \ref{subsec:3d_method}. In general, this method tends to underestimate FUV fluxes. 
\\
\\
Method 2: 2D Projected separation and distance uncertainty sampling. For the close pairs disc-massive star members of the same star-forming region we evaluated the FUV flux using their (2D) projected separation, which results in a general overestimate of the actual flux. More details on this approach are provided in in Sec. \ref{subsec:2d+3d_method}.
\\
\\
Method 3: Local density function and distance uncertainty sampling. we assumed spatial isotropy across a star-forming region, and we used the 2D geometry of the region to make arguments on the most probable 3D separation between stars. Specifically, we define the probability of finding each pair of stars at a certain (3D) separation, given their (2D) projected distance on the sky plane, and we sampled from the corresponding cumulative distribution function to estimate the corresponding FUV flux, as detailed in Sec. \ref{subsec:2dstat_3d_method}. This approach provides our best estimate of the FUV flux and the associated uncertainty.

\subsection{Method 1: Distance Uncertainty Sampling}\label{subsec:3d_method}
To account for the uncertainty in the relative distance affecting the final FUV flux value, a standard approach involves using a Monte Carlo sampling of the distributions of distances from Earth (or parallax values) of the single stars. 
This method is commonly used in nearby regions and where parallaxes are known with relative small uncertainties.
The ALS III catalogue, as well as the catalogue of massive stars by \citet{Zari_catalogue}, provides the 50$^{\mathrm{th}}$, 16$^{\mathrm{th}}$, and 84$^{\mathrm{th}}$ percentiles of the posterior distribution of distances. For the other stars contained in the Gaia catalogue, the median of the geometric distance posterior \citep{Bailer-Jones_geodist_gaia}, with 16$^{\mathrm{th}}$ and 84$^{\mathrm{th}}$ percentiles can be accessed
and evaluated for the the remnant stars. 
Larger uncertainties are registered in particular for faint objects, often embedded in gas clouds making the detection more challenging (e.g. discs in the $\rho$ Oph region), and for very luminous stars (e.g. $\sigma$Ori or $\theta^{1}$C). 
For the objects lacking parallax measurements we assigned parallax and parallax error as the median values of the nearest neighbours members of the same star-forming region having the same spectral type.
The FUV flux at each disc position is evaluated by randomly assigning the radial distance of individual objects based on their provided distributions, and then repeating this process multiple times to determine the median, 16$^{\mathrm{th}}$, and 84$^{\mathrm{th}}$ percentiles of the FUV flux. We will show in Sec. \ref{sec:synthetic_clusters} and \ref{sec:results} that this method tends in general to underestimate the FUV flux, as the range of values that the 3D separation between objects can assume when randomly sampling from individual extended distributions of line-of-sight distances, is large and the probability to retrieve the correct separation is small. 
This method may yield unreliable FUV flux estimate in cases where
distance uncertainties are comparable or larger than the separation between OBA star and star hosting disc. In such scenario, even a small change in the relative separation between the disc and the massive star (due to large uncertainties in the line-of-sight position compared to the other two components on the sky plane) can lead to substantial variations in the final flux value. 

\subsection{Method 2: 2D Projected Separation and Distance Uncertainty Sampling }\label{subsec:2d+3d_method}
The highest contribution to the FUV flux at the position of a certain star-hosting disc is given by massive stars members of the same star-forming regions, as these are the closest to the object in exam. However, the flux induced by these stars is subject to large uncertainty due to the problem raised above, where a small uncertainty in the line-of-sight position of two close objects can propagate in a large uncertainty in the FUV flux estimate.
Therefore, for each star-forming region we focused on these massive stars defined as members of region. 
Assuming spatial isotropy within a given region, all the directions have the same properties, meaning that stars with closer projected separations are more likely to be physically close (i.e. in 3D space). Hence, an upper limit of the FUV flux induced by massive stars members of a certain region can be obtained considering the separation between them and the star-hosting discs (members of the same region) projected on the sky plane at a line-of-sight distance corresponding to the object presenting the smallest distance uncertainty. 
For the massive stars outside the a certain region, we applied Method 1, described in the previous Section, where an estimate of the uncertainty on the FUV flux is provided by randomly sampling the distribution of line-of-sight distances of single objects. 
While this approach offers an initial estimate of the flux when parallax measurements are unavailable, it has limitations. Specifically, as the primary source of flux uncertainty for discs arises from the nearest massive stars, for which we only consider 2D separation and ignore errors (even when available) in the third dimension, we lose information about the accuracy of the flux calculation.
Indeed, the only source of flux uncertainty in this method is produced by the separation between discs and the massive stars located outside their star-forming region, which is small as more distant stars influence less the flux calculation.

\subsection{Method 3: Local Density Function and Distance Uncertainty Sampling}\label{subsec:2dstat_3d_method}
For the massive stars and star-hosting discs that are member of the same star-forming region, which can present errors on their 3D relative distance comparable or larger than their 3D separation, we can improve Method 2 (Sec. \ref{subsec:2d+3d_method}). We can do this by appealing to the geometry of the region to find the best estimate of the 3D separation between stars given their 2D relative distance. 
We assume that (i) the stellar members of a given star-forming region have well-defined separations on the sky plane, meaning that we neglect errors in RA and Dec positions, as these are minimal compared to the separations between stars and are smaller than the uncertainties in parallax, and (ii) spatial isotropy applies throughout the region.
Considering a stellar member $i$, we employ the Bayes theorem to calculate the differential probability to find a second member $j$ at a certain 3D separation $r_{ij}$, assuming to know the 2D separation $R_{ij}$:
\begin{equation}
    \mathrm{d} P (r_{ij} | R_{ij}) = \frac{ \mathrm{d} P (R_{ij} | r_{ij}) \ \mathrm{d} P(r_{ij}) }{ \mathrm{d} P(R_{ij}) },
    \label{eq:Bayes_rR}
\end{equation}
In order to derive an analytic expression for Eq. \eqref{eq:Bayes_rR} (i.e. likelihood and prior), we start defining the differential probability to find a star $j$ at projected separation $R_{\mathrm{ij}}$ from a star $i$, in a small area $\mathrm{d} A = 2\pi R_{ij} \mathrm{d} R$:
\begin{equation}
    \mathrm{d}P (R_{ij}) = \hat{\Sigma}_{\mathrm{pairs}} (R_{ij}) \ \mathrm{d} A,
    \label{eq:prob_sigma}
\end{equation}
where $\hat{\Sigma}_{\rm{pairs}} (R)$ is the normalised probability density function of all the projected separations between the members of the star-forming region.
Under the assumption of isotropy, the Abel's theorem \citep{Abel_1826} states that considering the (2D) surface density of neighbouring stars at distance $R$, $\hat{\Sigma}_{\mathrm{pairs}} (R)$, in the form 
\begin{equation}
    \hat{\Sigma}_{\mathrm{pairs}} (R) = 2 \int_{R}^{\infty} \hat{\rho}_{\mathrm{pairs}} ( r) \frac{R}{\sqrt{r^{2} - R^{2}}} \mathrm{d}r,
    \label{eq:Abel_sigma}
\end{equation}
the corresponding (3D) volume density, $\hat{\rho}_{\mathrm{pairs}} ( r)$, is uniquely defined as:
\begin{equation}
    \hat{\rho}_{\mathrm{pairs}} (r) = - \frac{1}{\pi} \int_{r}^{\infty} \frac{ \mathrm{d} \hat{\Sigma}_{\mathrm{pairs}} (R) }{ \mathrm{d} R} \frac{1}{\sqrt{ r^{2} - R^{2}}} \mathrm{d} R ,
    \label{eq:Abel_rho}
\end{equation}
which is valid if $\hat{\rho}_{\mathrm{pairs}} (r) \rightarrow 0$ as $r \rightarrow \infty$. As Abel's inversion requires spherical symmetry to project a surface density in a volume density, we applied the theorem to the distribution of separations between pairs of stars in the isotropic case, and to the separations from the centre of the region for a centrally concentrated region.
Using Eq. \ref{eq:Abel_rho}, we can rewrite Eq. \ref{eq:Bayes_rR} for a single pair of stars $ij$ as:
\begin{equation}
    \mathrm{d} P (r_{ij} | R_{ij}) = \mathrm{d} P (R_{ij} | r_{ij}) \ \mathrm{d} P(r_{ij}) = \begin{cases} \frac{ 4 \pi R_{ij} r_{ij}}{\sqrt{r_{ij}^2 - R_{ij}^2}}\hat{\rho}_\mathrm{pairs} ( r_{ij})  & R_{ij} < r_{ij} \\ 0 & \mathrm{otherwise}
    \end{cases},
    \label{eq:prob_rR_final}
\end{equation}
which defines our Probability Distribution Function (PDF), and where the factor multiplying the volume density is included to account for the geometrical effect when passing from 2D to 3D geometry.
Using the precise RA and Dec measurements provided by Gaia DR3 and Hipparcos, we define the separation $R$ as the angular separation between two stars on the celestial sphere (evaluated using the Vincenty formula \citeauthor{Vincenty_formula} \citeyear{Vincenty_formula}) projected to the median distance of the considered region. For each star-forming region included in this work, we used membership census available in literature (see Sec. \ref{sec:results} for references of the census used) to evaluate $\hat{\Sigma}_{\mathrm{pairs}}(R)$ (hereafter $\hat{\Sigma}$) as the distribution of the projected separations between pairs of stars in a same region. Subsequently, we estimate the 3D density profile $\hat{\rho}_{\mathrm{pairs}}(r)$ (hereafter $\hat{\rho}$) using Eq. \eqref{eq:Abel_rho}, where $r$ is the 3D (spherical) separation between stars. 
Finally, we sample the probability distribution defined in Eq. \eqref{eq:prob_rR_final} to derive the 3D separation between close stellar pairs, and we use this parameter to evaluate the FUV flux and the corresponding uncertainty.

In cases where the geometry of the region is nearly centrally concentrated, we can assume spherical isotropy across the entire region and still use the expression in Eq. \eqref{eq:prob_rR_final}, but considering $r_{ij}$ and $R_{ij}$ as the 3D and 2D separations from the centre of the distribution (i.e. $r_{ij}$ is the radius of the sphere centred in the centre of the distribution, and $R_{ij}$ is radius in polar coordinates when neglecting the line-of-sight component).
This method can be applied in regions like the ONC and $\sigma$ Ori, where stellar members are distributed almost symmetrically around the most luminous star, which in these cases are $\theta^1$C and $\sigma$ Ori, respectively.

\subsection{Testing on Synthetic Stellar Clusters the Three Methods for Evaluating the FUV Flux}\label{sec:synthetic_clusters}
The three approaches to evaluate the FUV flux, described in the previous Section, and in particular the effectiveness of using the local density function to infer the 3D separation between stars in a same star-forming region, are examined through synthetic stellar clusters.
\begin{figure}[htbp]
    \centering
    
    \begin{minipage}[b]{0.45\textwidth}
        \centering
        \includegraphics[width=\textwidth]{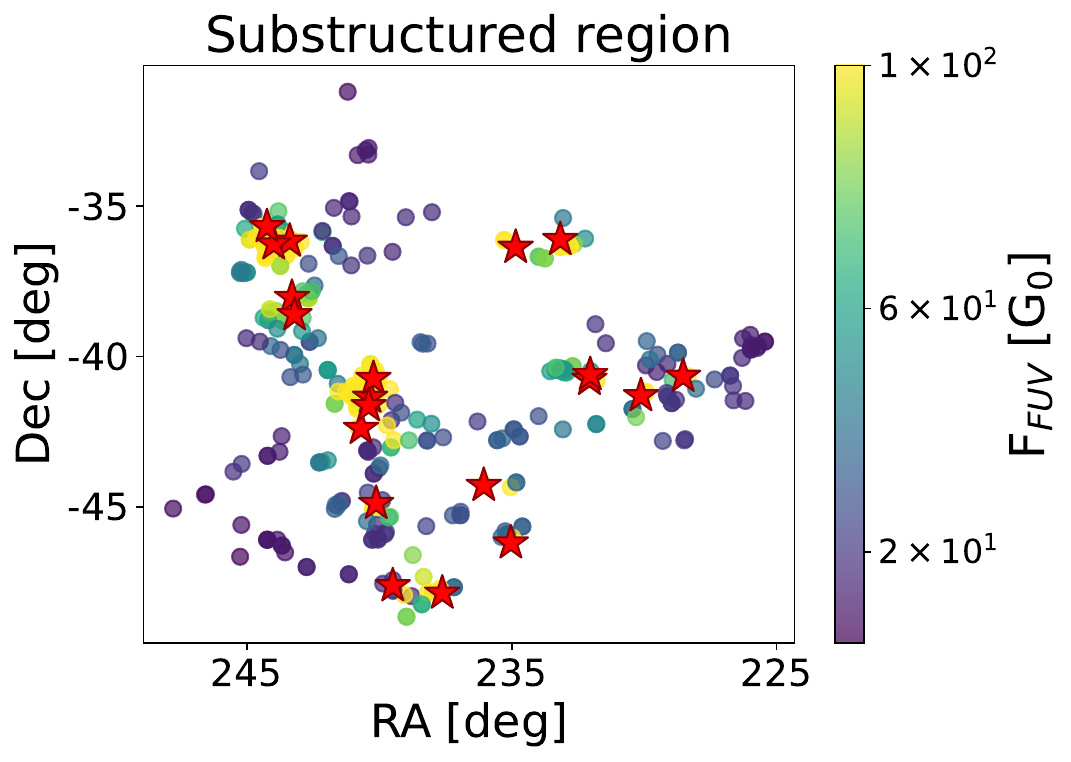}
    \end{minipage}
    \begin{minipage}[b]{0.45\textwidth}
        \centering
        \includegraphics[width=\textwidth]{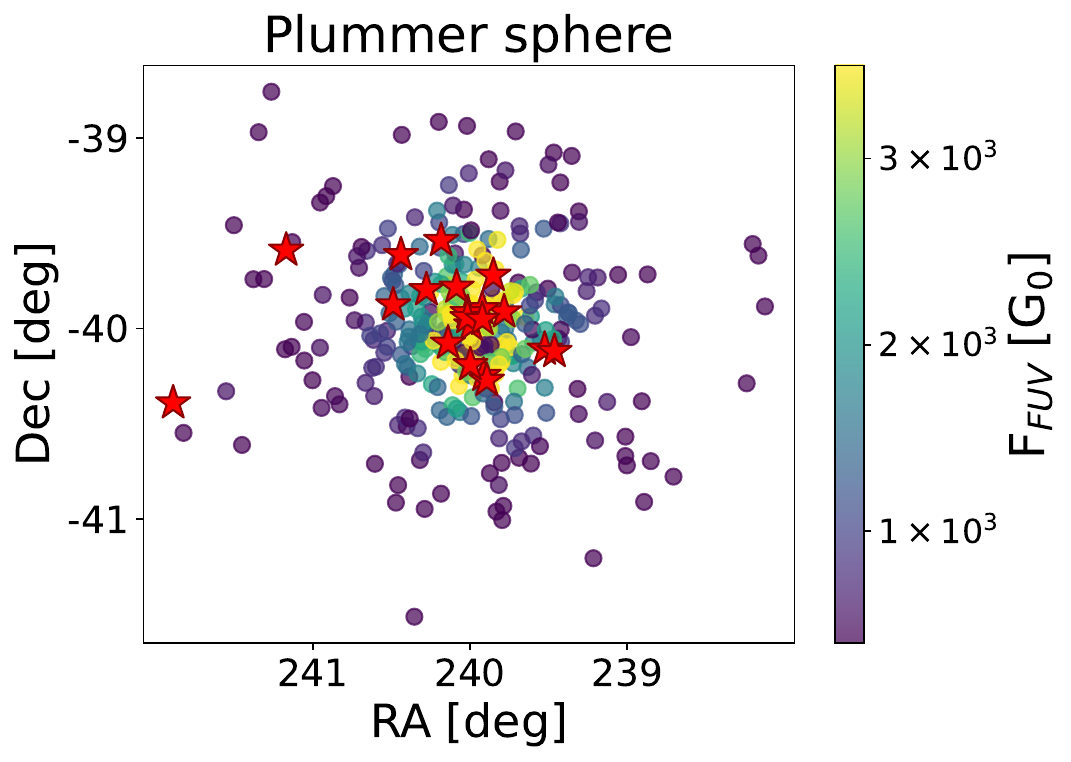}
    \end{minipage}
    
    \caption{Position of the low-mass and high-mass stars in the synthetic stellar clusters: substructured region in the top panel and Plummer sphere in the bottom panel.}
    \label{fig:2d_maps_synthetic}
\end{figure}
\begin{figure}[htbp]
    \centering
    
    \begin{minipage}[b]{0.4\textwidth}
        \centering
        \includegraphics[width=\textwidth]{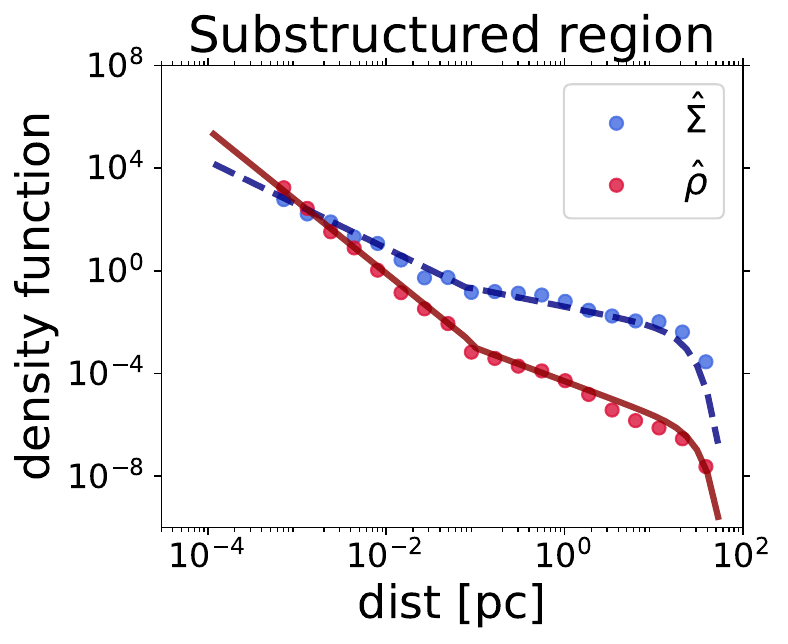}
    \end{minipage}
    \begin{minipage}[b]{0.4\textwidth}
        \centering
        \includegraphics[width=\textwidth]{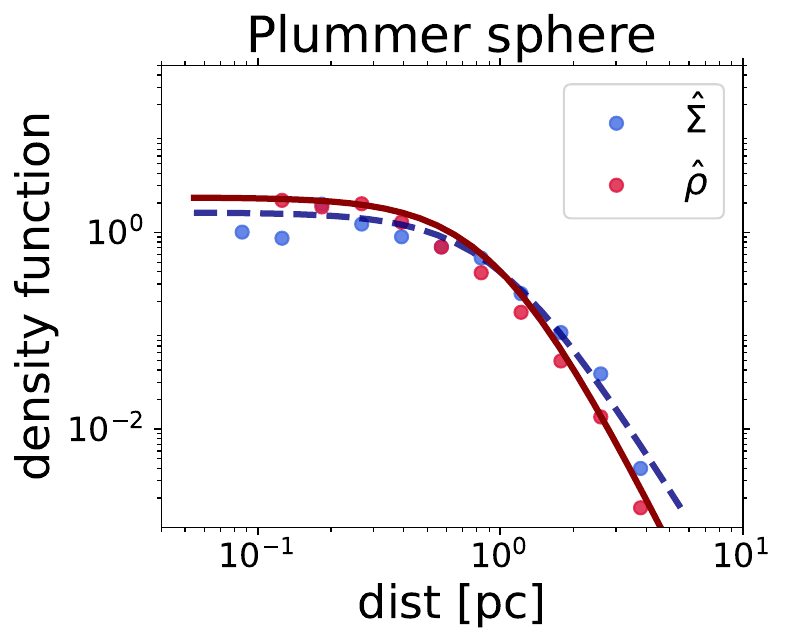}
    \end{minipage}
    
    \caption{Surface density ($\hat{\Sigma}$) and volume density ($\hat{\rho}$) profiles related to the simulated clusters. The x-axis represents the separation between pairs of stars in the substructured region (top panel), and the distance from the cluster centre in the Plummer sphere case (bottom panel).}
    \label{fig:density_profiles_synthetic}
\end{figure}
\begin{figure}[htbp]
    \centering
    
    \begin{minipage}[b]{0.4\textwidth}
        \centering
        \includegraphics[width=\textwidth]{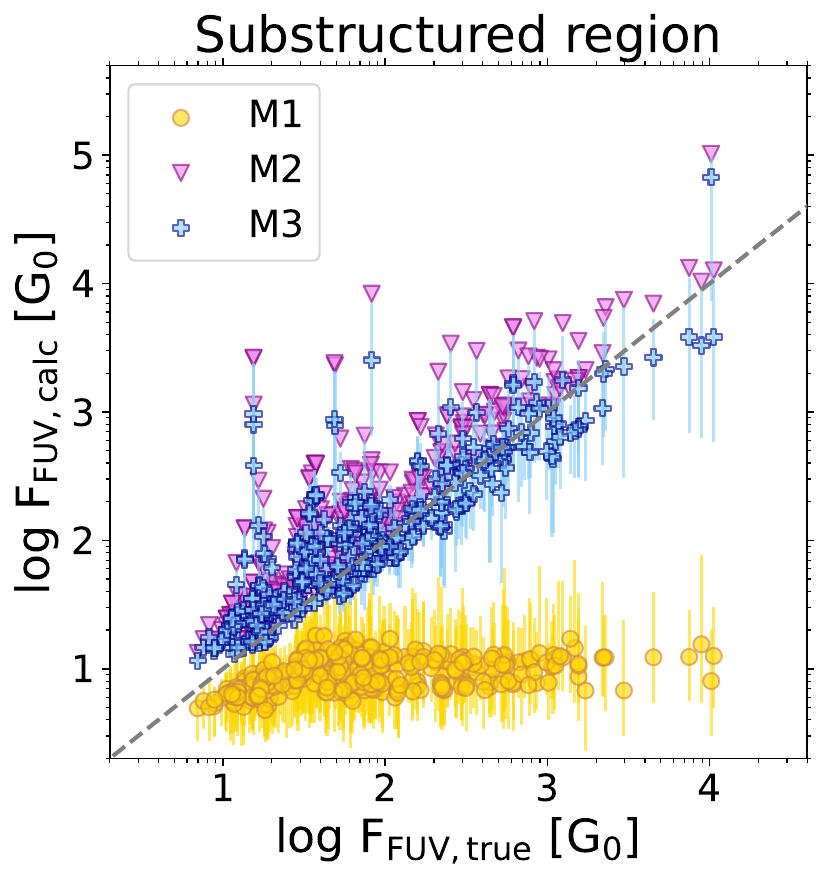}
    \end{minipage}
    \begin{minipage}[b]{0.4\textwidth}
        \centering
        \includegraphics[width=\textwidth]{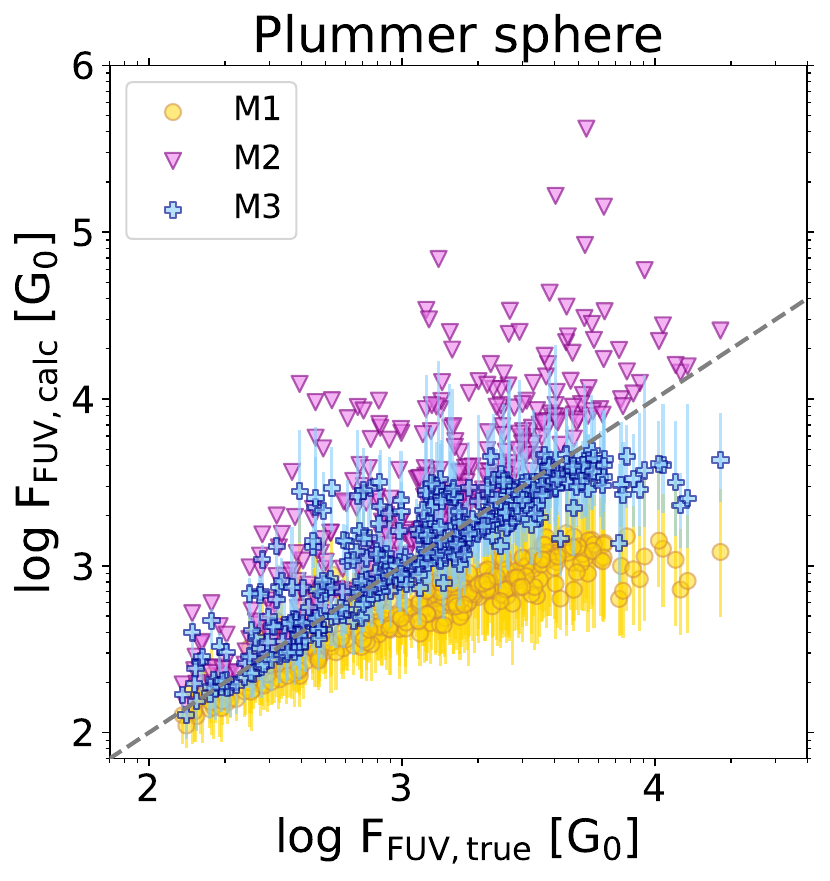}
    \end{minipage}
    \caption{Comparison between the true FUV flux experienced by the low-mass stars, as evaluated from the simulated clusters, and the value resulting from applying the three approaches described in Sec. \ref{subsec:3d_method} (M1), \ref{subsec:2d+3d_method} (M2), and \ref{subsec:2dstat_3d_method} (M3).}
    \label{fig:fuv_synthetic}
\end{figure}
In order to highlight how the density profile depends on the geometry of the star-forming region, we simulated two regions with identical numbers of high-mass (20) and low-mass (340) stars, but differing geometries: the first presenting internal substructures, and the second modelled as a Plummer sphere (to resemble a spherically symmetric cluster), shown respectively in top and bottom panel of Fig. \ref{fig:2d_maps_synthetic}.
The substructured synthetic cluster is based on a random subset from the spatial distribution of stars in \citet{Winter_running_with_bulls}, considering the simulation snapshot at 1 Myr.
Assuming isotropy, the density profile associated to the substructured region depends on the separation between pairs of stars.
Conversely, under the condition of spherical symmetry around the centre of the region, as in the Plummer sphere case, the density profile is a function of the distance from the centre.
The surface and volume density profile of the two simulated regions are shown in Fig. \ref{fig:density_profiles_synthetic}. For the substructured region, $\hat{\rho}$, which consists of the Abel transform of $\hat{\Sigma}$, can be analytically prescribed by a double power law exponentially truncated:
\begin{equation}
    \hat{\rho}(r) =
    \Bigg\{
    \begin{array}{ll}
        \alpha_{1} \ r^{-\beta_{1}}, & r \leq r_{\mathrm{lim}}, \\
        \alpha_{2} \ r^{-\beta_{2}} \ \exp{ [ -(r/r_{\mathrm{b}})^{-\gamma} ] }, &  r > r_{\mathrm{lim}},
    \end{array}
\end{equation}
where $r$ is the separation between pairs of stars in 3D space (spherical coordinate), normalized to 1 pc in our expressions, and the parameters adopted in the presented case are $\alpha_{1} = 2\times10^{-6}$, $\beta_{1} = 2.8$, $\alpha_{2} = 5\times10^{-5}$, $\beta_{2} = 1.3$, the length-scale of the cluster $r_{\mathrm{b}}=25$ pc (which set the boundary of the cluster), $\gamma = 2.6$, and the limit separation $r_{\mathrm{lim}} = 0.1$ pc, which sets the change of power law steepness. These parameters vary based on the specific geometry of the region.
The Plummer volume density profile used is:
\begin{equation}
    \hat{\rho}(r) = \alpha \ (1 + r^{2})^{-5/2},
\end{equation}
where $r$ is the distance from the centre of the cluster in 3D space, and we assumed ad Plummer radius of 1 pc and $\alpha \simeq 2.4$.
The density profiles adopted for the various star-forming regions investigated in this work are summarised in Appendix \ref{Appendix:density_profiles}. 

We evaluated the FUV flux experienced by the low-mass stars due to the irradiation of the high-mass stars, using the three approaches described in Sec. \ref{subsec:3d_method}, \ref{subsec:2d+3d_method}, and \ref{subsec:2dstat_3d_method}. In particular, in order to apply the distance sampling method, we introduced an error on the distance of the stars comparable to the length scale of the region ($\sim$25 pc for the substructured region, and $\sim$1 pc for the Plummer sphere), which is relatively large compared to the median parallax error in the most nearby regions (e.g. Taurus, Lupus), but is reasonable in more distant regions (e.g. Orion).
In Fig. \ref{fig:fuv_synthetic} the FUV fluxes resulting from using the three approaches are compared with the true values. 
Method 1 (Sec \ref{subsec:3d_method}), shown in yellow in Fig. \ref{fig:fuv_synthetic}, tends to significantly underestimate the actual flux, of orders of magnitude for some stars.
This result highlights the error in FUV flux that can result from sampling Gaia parallaxes and parallax uncertainties, and the power of our proposed approach.
The use of Method 2 (Sec. \ref{subsec:2d+3d_method}) leads to a general overestimate of the flux. Specifically, the fraction of total FUV fluxes reproduced (within 1$\sigma$) by Method 1 is 1.3\% for the substructured cluster and 21\% for the Plummer sphere, while Method 2 reproduces the $\sim$7.5\% of the true values in both cases.
Employing the local density function (Method 3) yields FUV fluxes that are consistent with the true values within 1$\sigma$ error bars for 48$\%$ of the low-mass stars in the substructured region, and 68$\%$ of the low-mass stars in the Plummer sphere case. 
In the case of the substructured region, we obtained a smaller fraction of successfully reproduced fluxes with respect to the Plummer sphere. This is caused by the fact that we neglected the correlation between stellar pairs when employing the Bayes theorem. Specifically, in our approach, the 3D separation of each pair of low-mass - high-mass star is independently sampled from a cumulative distribution function (CDF). However, the separation between each pair of stars is not completely independent from the others and a covariance term should be included in the analytic formulation.
Not accounting for this term results in a general underestimate of the flux uncertainty.
However, the uncertainty in the flux is more underestimated as the number of massive stars contributing to the flux rises. Therefore, if a star-forming region hosts a small number of massive stars, the error caused by neglecting the covariance term is relatively small.
Accounting for this term requires a more detailed analytic work, and further testing on simulated clusters. As the resulting fluxes are reproducing within the error bars a substancially larger fraction of true values, this study is left to future investigation.   
Conversely, in the Plummer sphere, as all the separations are defined with respect to the fixed centre of the cluster, we did not investigate each stellar pair independently, but the entire cluster: we sample from a CDF the 3D separation of all the stars from the centre, generating entire random clusters, and then we evaluate the FUV flux within each cluster.

The fact that Method 3 (Sec. \ref{subsec:2dstat_3d_method}) yields more than twice the percentage of FUV fluxes consistent with true values compared to other methods, highlights the strength and advantage of our proposed approach over the investigated alternative methods.
\section{Results: a catalogue of FUV flux values for nearby discs}\label{sec:results}
\begin{figure*}[htbp]
    \centering  
    \begin{minipage}[b]{0.4\textwidth}
        \centering
        \includegraphics[width=\textwidth]{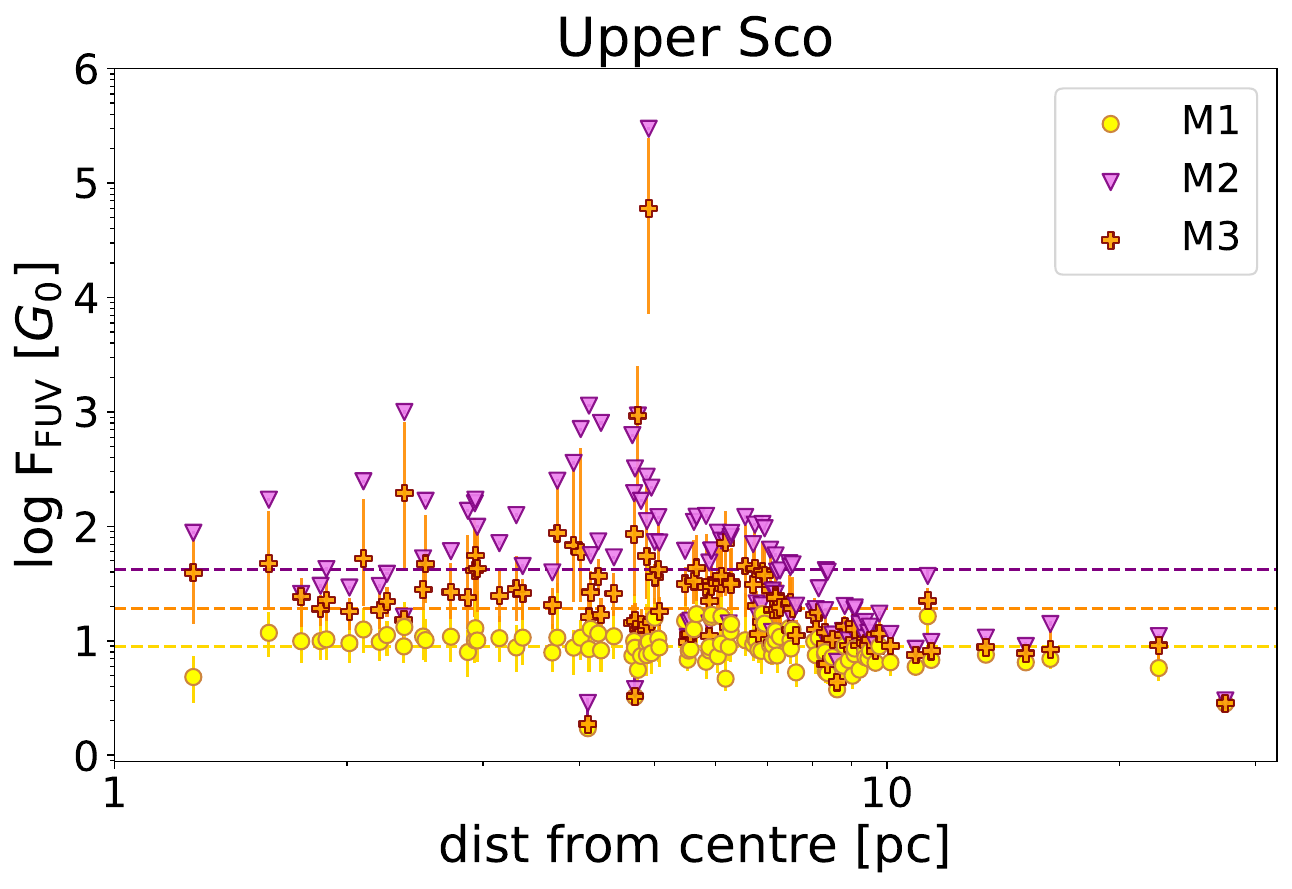}
    \end{minipage}
    \hspace{0.04\textwidth}
    \begin{minipage}[b]{0.4\textwidth}
        \centering
        \includegraphics[width=\textwidth]{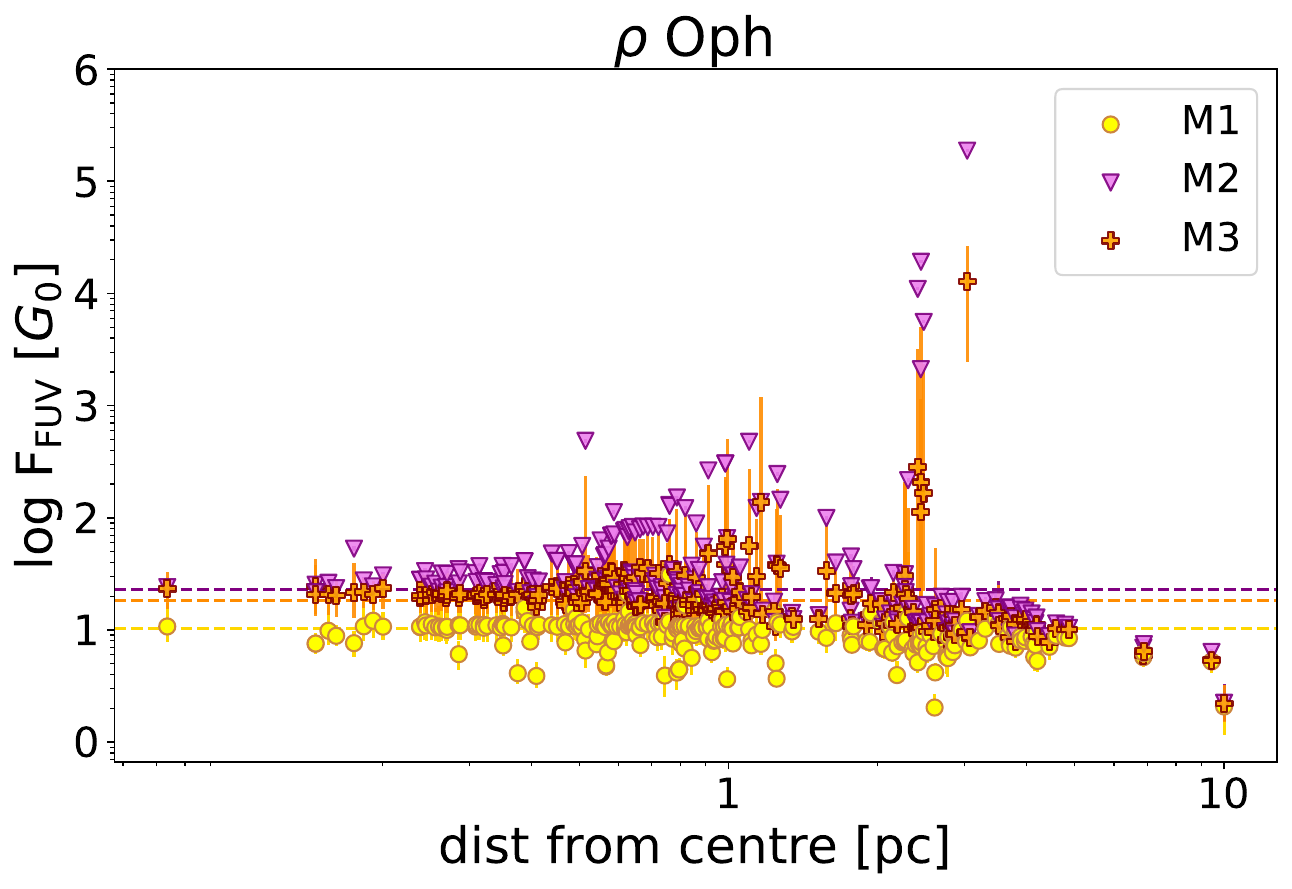}
    \end{minipage}
    \begin{minipage}[b]{0.4\textwidth}
        \centering
        \includegraphics[width=\textwidth]{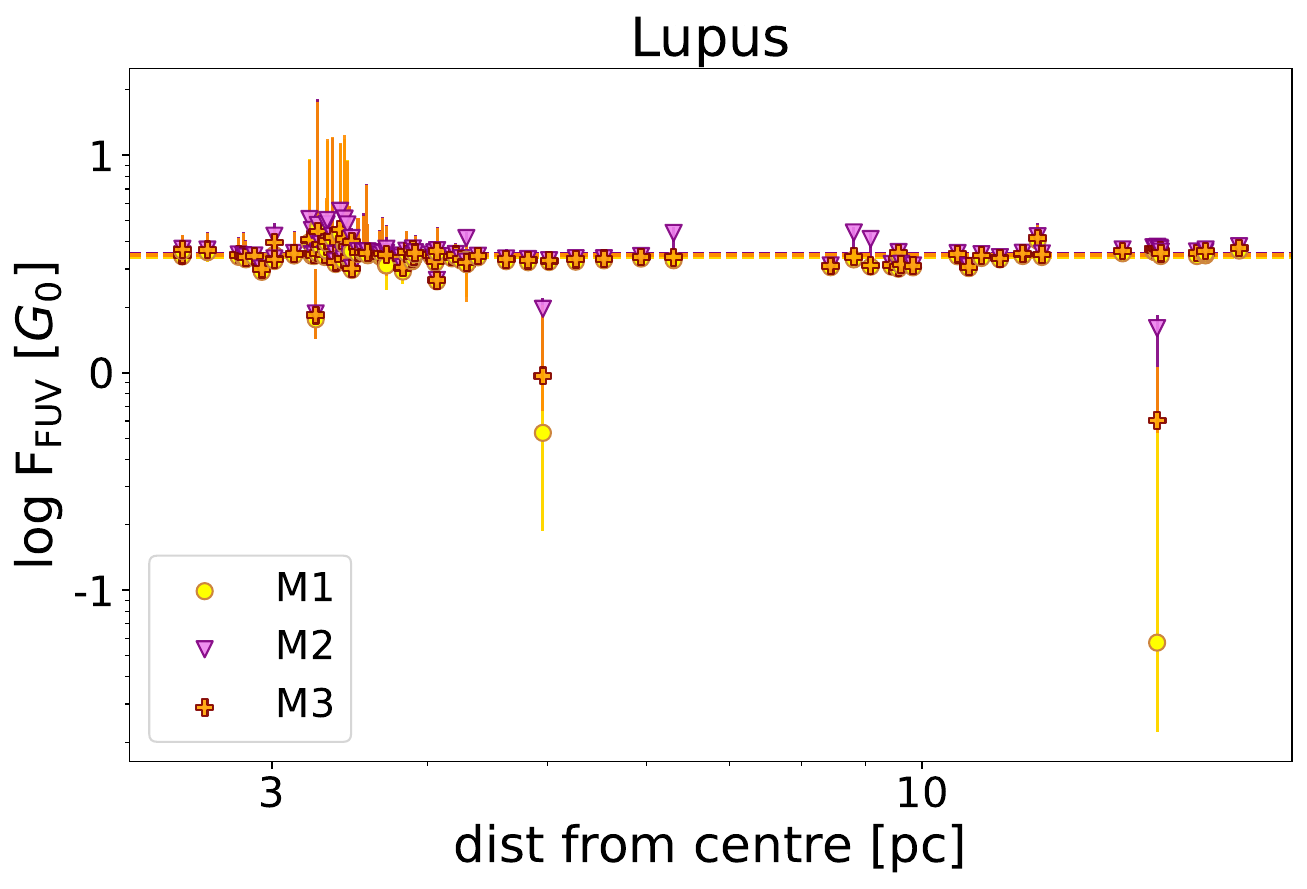}
    \end{minipage}
    \hspace{0.04\textwidth}
    \begin{minipage}[b]{0.4\textwidth}
        \centering
        \includegraphics[width=\textwidth]{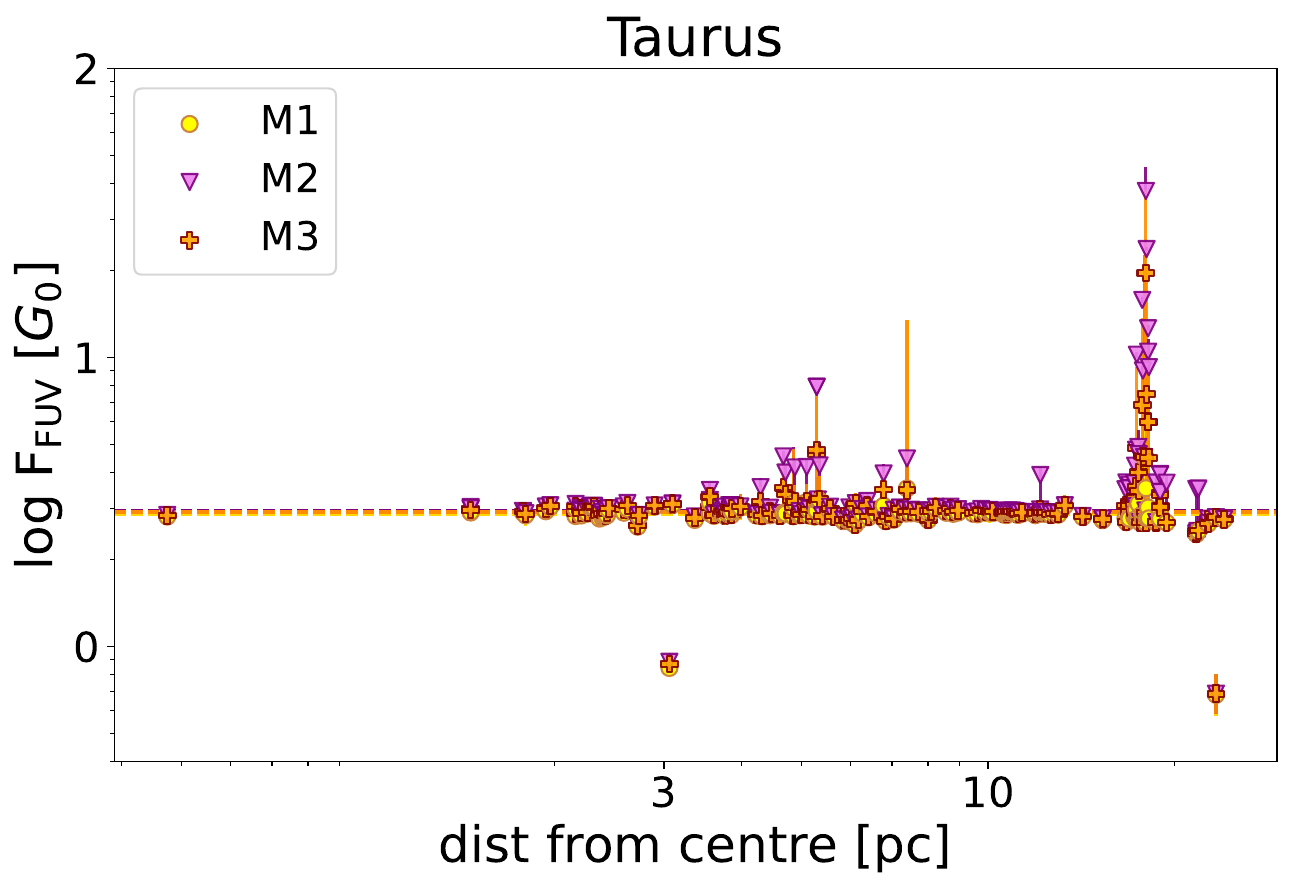}
    \end{minipage}
    \begin{minipage}[b]{0.4\textwidth}
        \centering
        \includegraphics[width=\textwidth]{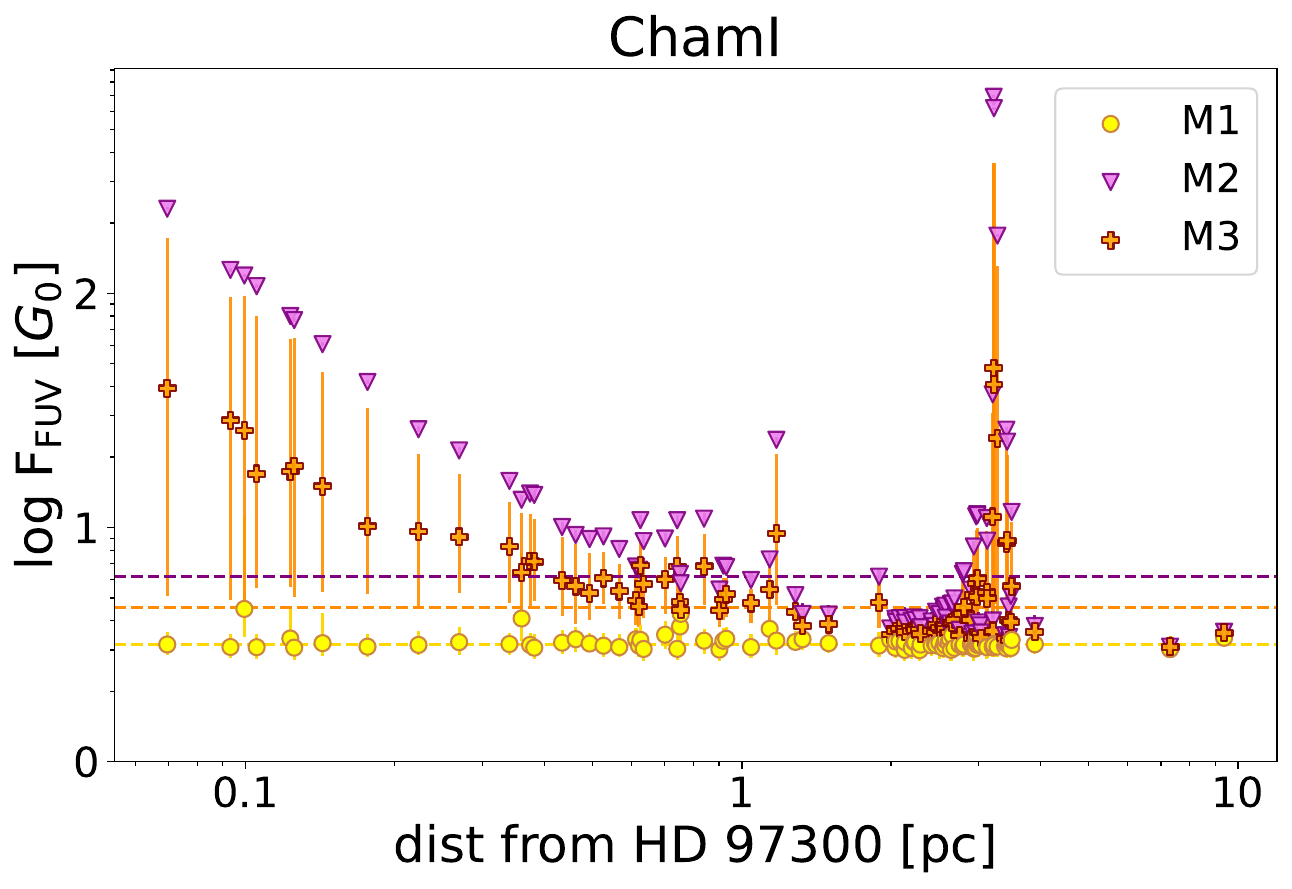}
    \end{minipage}
    \hspace{0.04\textwidth}
    \begin{minipage}[b]{0.4\textwidth}
        \centering
        \includegraphics[width=\textwidth]{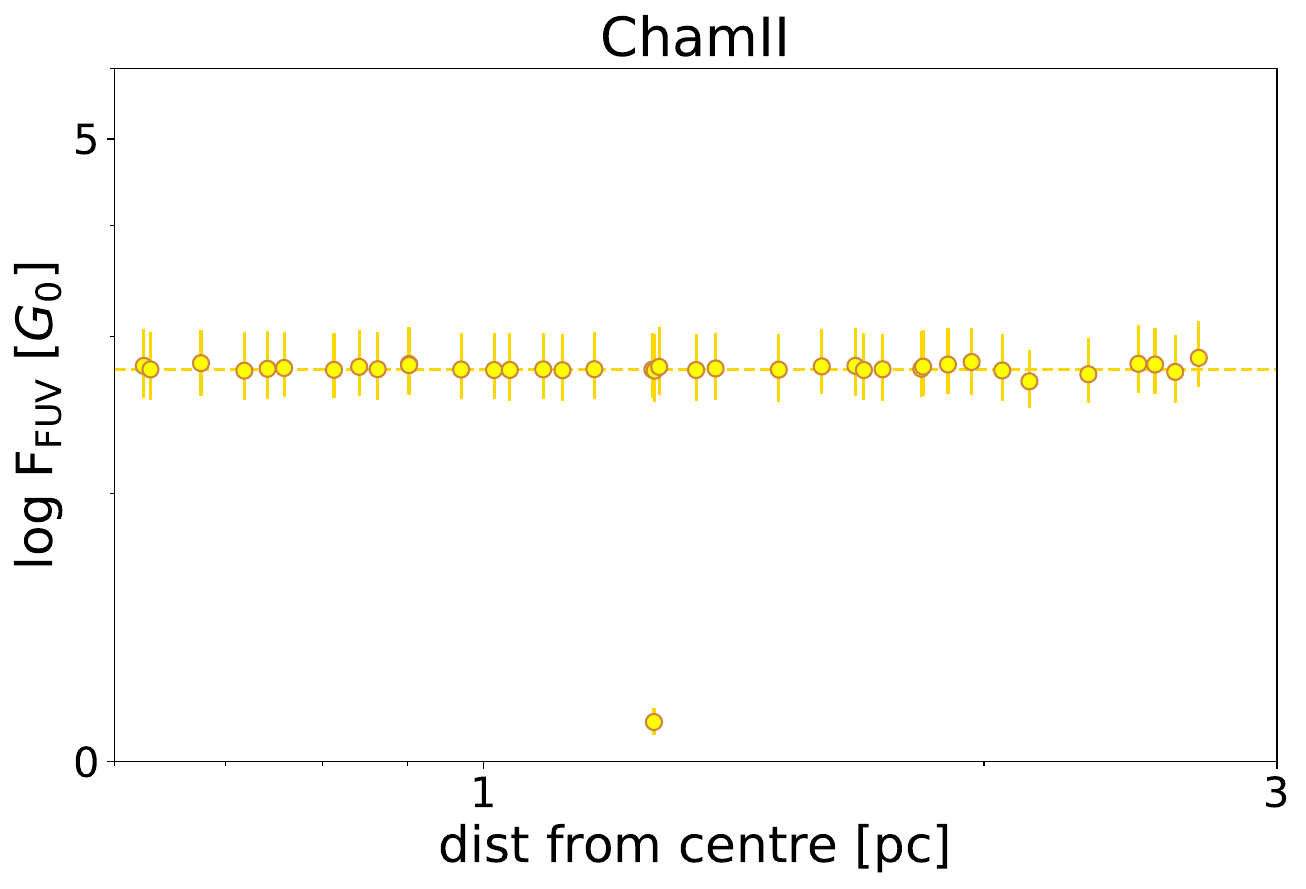}
    \end{minipage}
    \begin{minipage}[b]{0.4\textwidth}
        \centering
        \includegraphics[width=\textwidth]{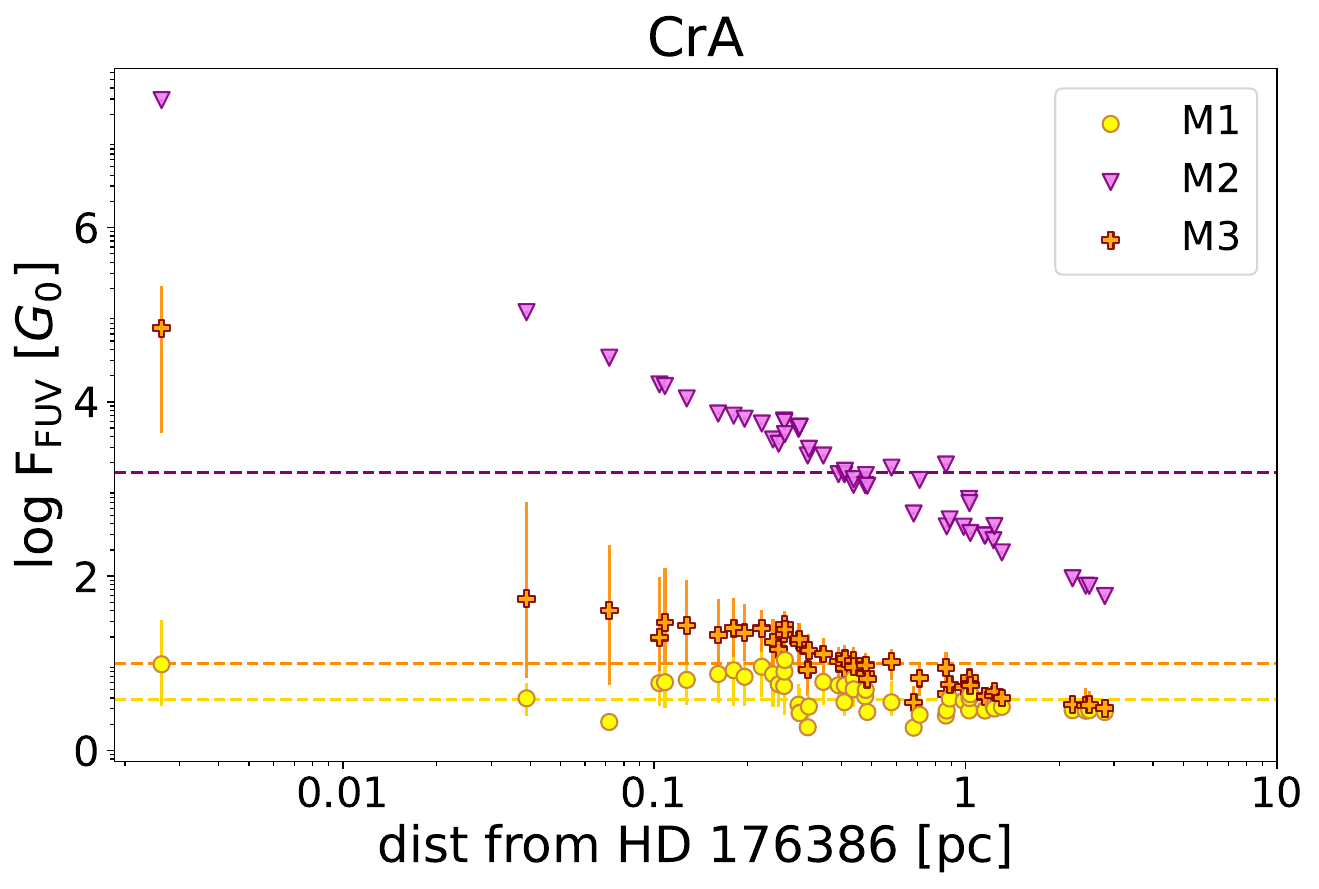}
    \end{minipage}
    \caption{FUV flux experienced by discs in seven regions located at $\lesssim$200 pc from the Sun. The three approaches used to evaluate the FUV flux are presented in yellow (Method 1), purple (Method 2), and orange (Method 3). The dashed lines show the median FUV flux value for each calculation method and region. x-axis: distance from the most massive star, or the centre of the region (when the distribution of OBA stars, and therefore FUV fluxes, is $\sim$ uniform across the region). }
    \label{fig:nearby_fuvs}
    \end{figure*}
\begin{figure*}[htbp]
    \centering
    \begin{minipage}[b]{0.4\textwidth}
        \centering
        \includegraphics[width=\textwidth]{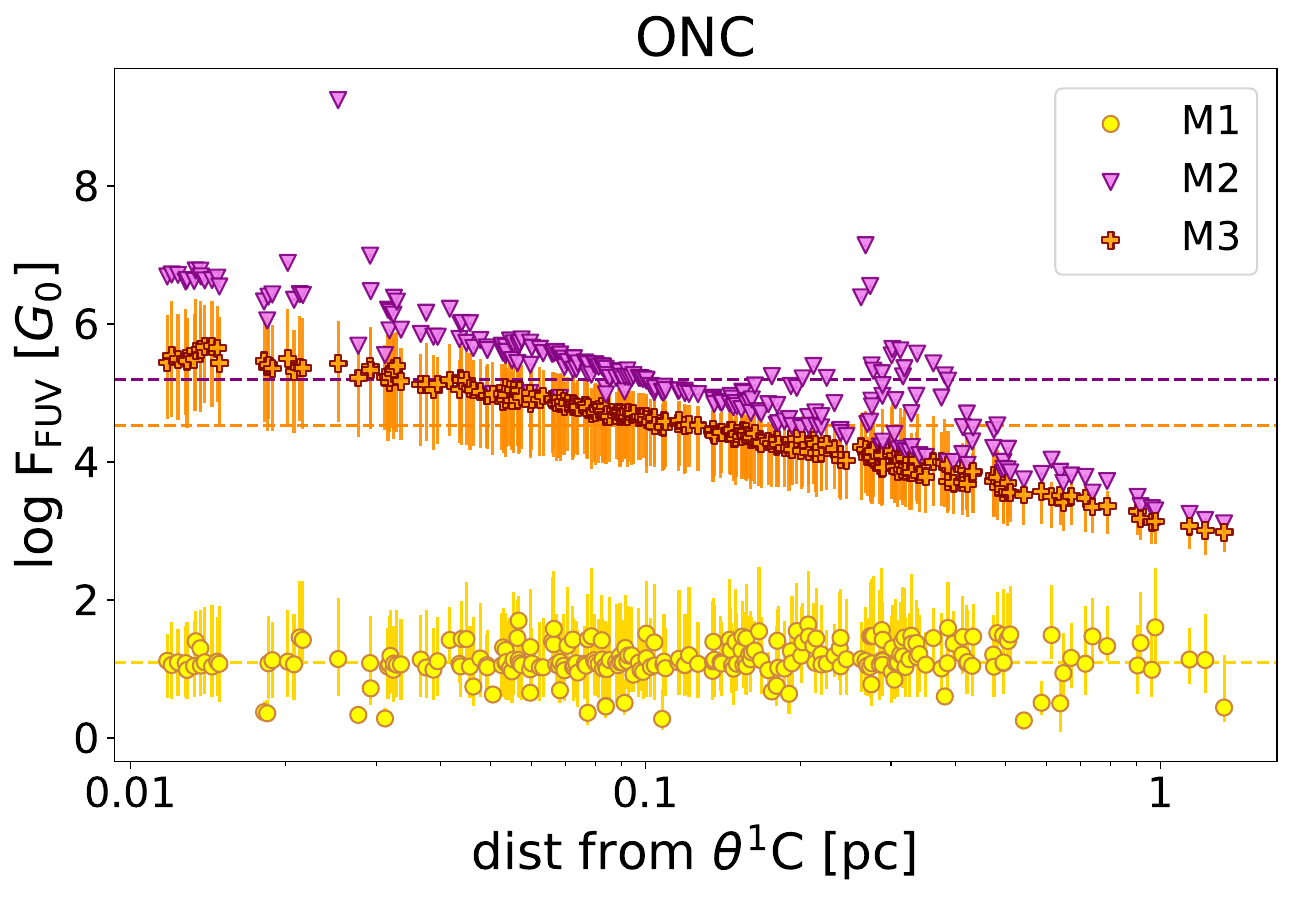}
    \end{minipage}
    \hspace{0.04\textwidth}
    \begin{minipage}[b]{0.4\textwidth}
        \centering
        \includegraphics[width=\textwidth]{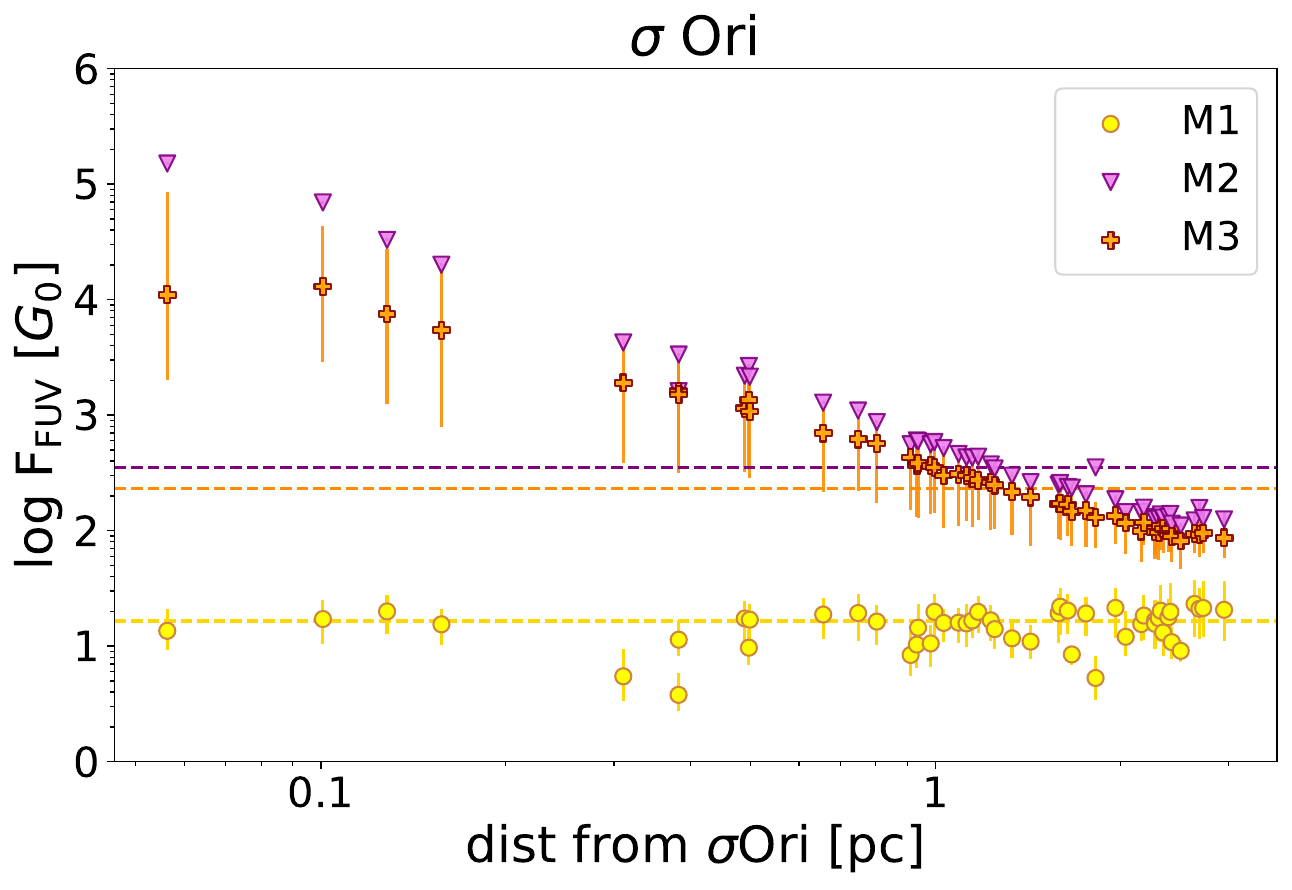}
    \end{minipage}
    \begin{minipage}[b]{0.4\textwidth}
        \centering
        \includegraphics[width=\textwidth]{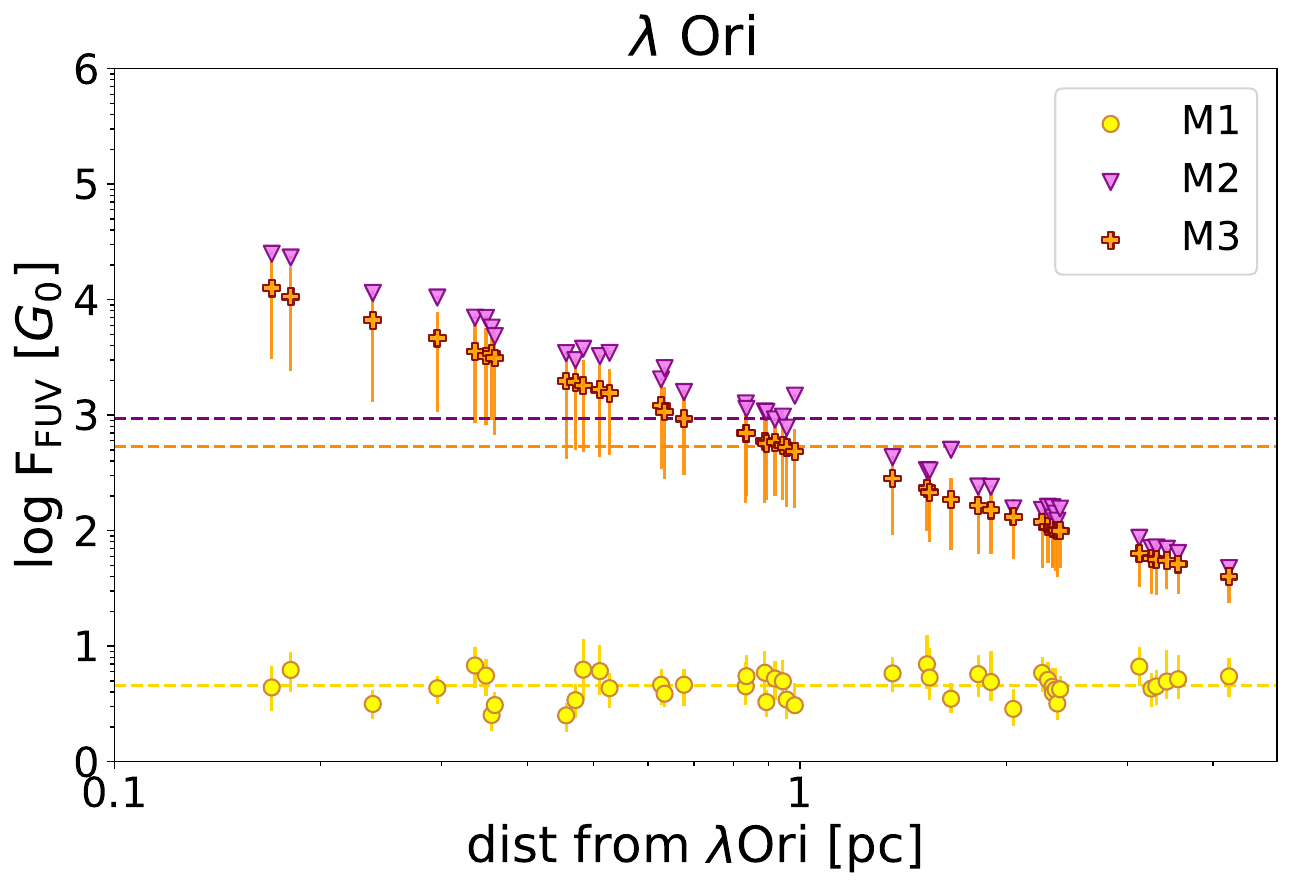}
    \end{minipage}
    \hspace{0.04\textwidth}
    \begin{minipage}[b]{0.4\textwidth}
        \centering
        \includegraphics[width=\textwidth]{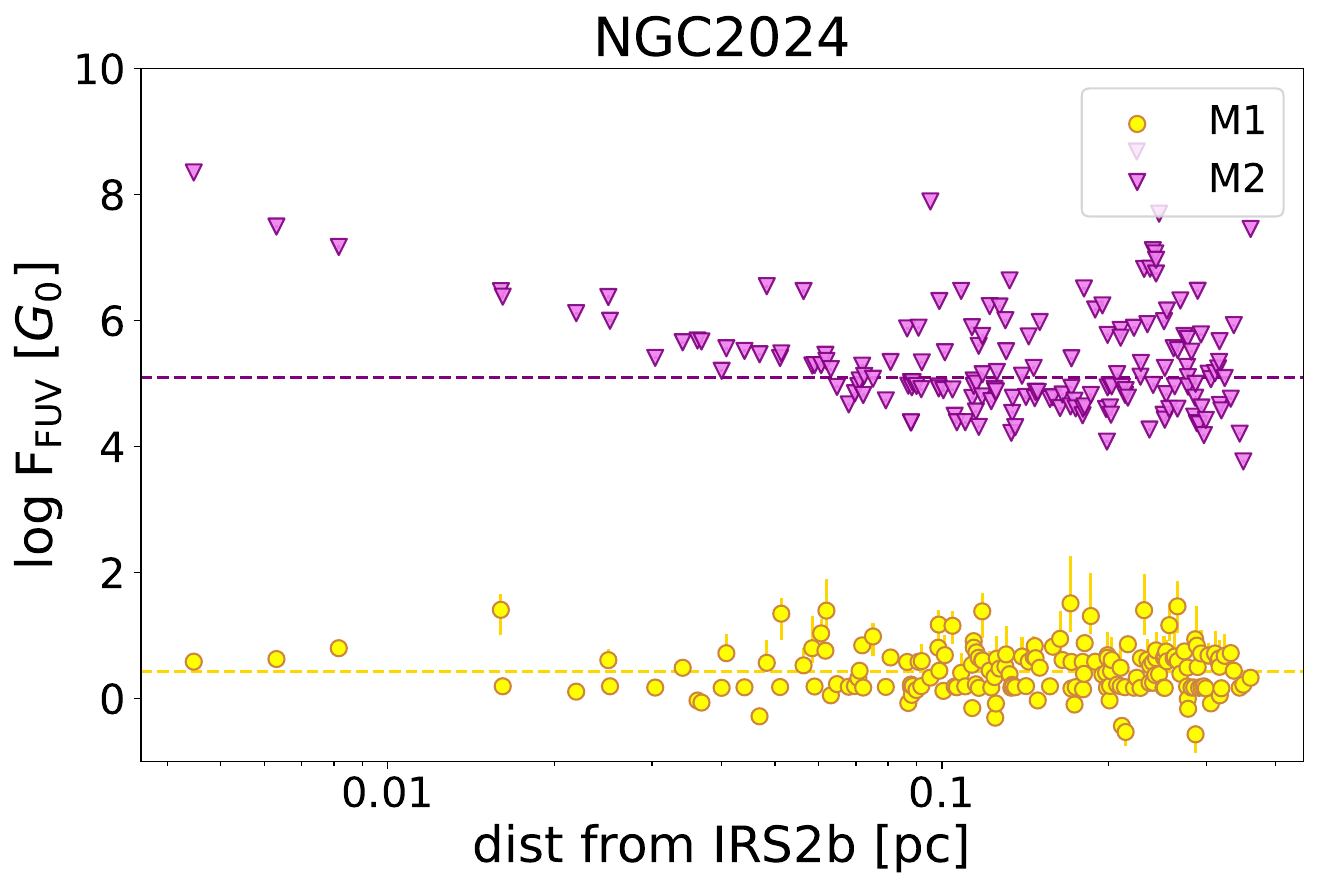}
    \end{minipage}
    \begin{minipage}[b]{0.4\textwidth}
        \centering
        \includegraphics[width=\textwidth]{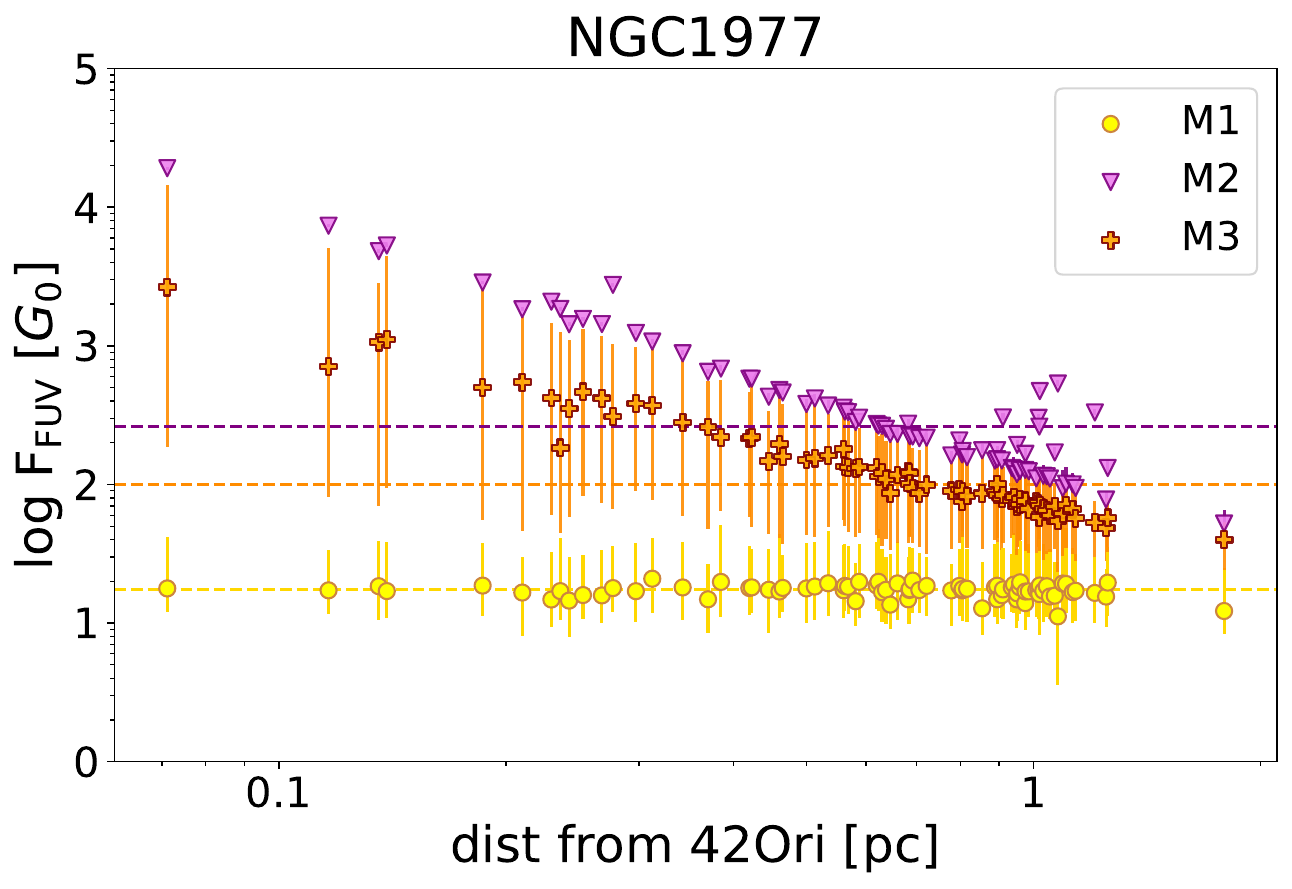}
    \end{minipage}
    \hspace{0.04\textwidth}
    \begin{minipage}[b]{0.4\textwidth}
        \centering
        \includegraphics[width=\textwidth]{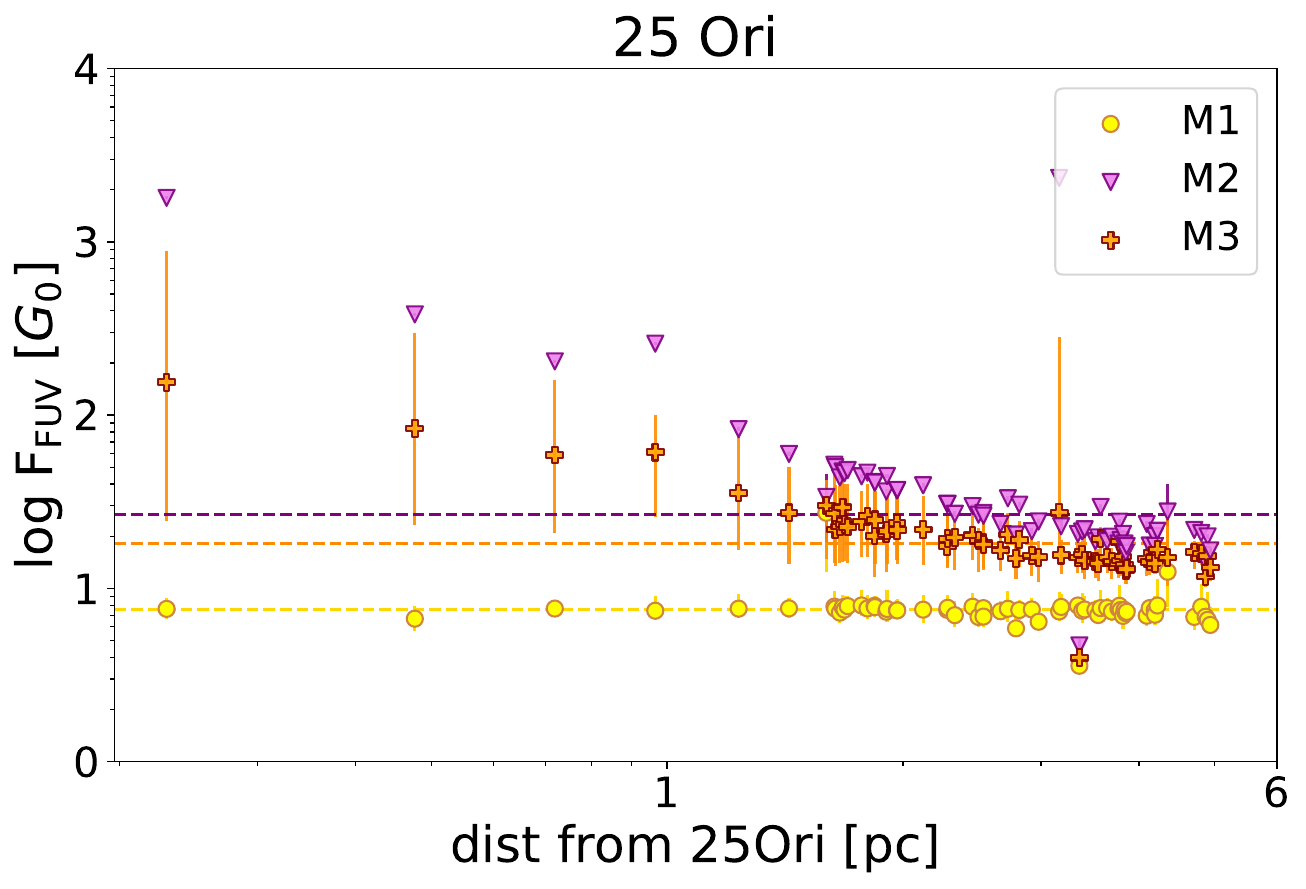}
    \end{minipage}
    \caption{FUV fluxes experienced by discs in six star-forming regions in Orion. The different colours correspond to different methods for the calculation of the FUV flux, as introduced in Sec. \ref{sec:fuv_calculation_methods_dist_eval}. The dashed lines show median FUV flux values for each region, and for calculation method used. x-axis contains the distance from the most massive star in the region. In NGC 2024 the local density distribution method cannot be applied (and therefore orange points are not contained in that panel), as an accurate density profile of the region cannot be currently retrieved.}
    \label{fig:orion_fuvs}
    \end{figure*}
\begin{figure*}[htbp]
    \centering
    \begin{minipage}[b]{0.4\textwidth}
        \centering
        \includegraphics[width=\textwidth]{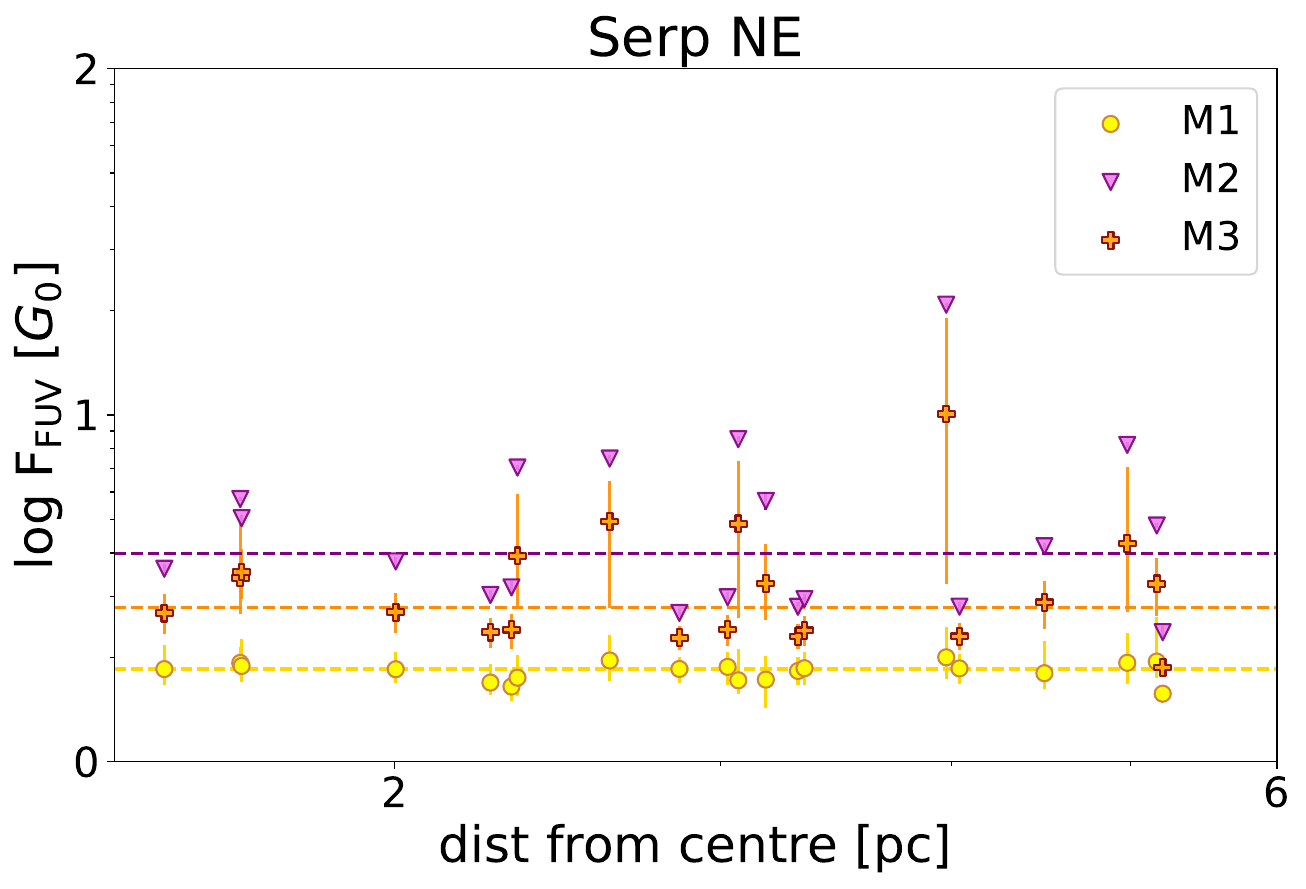}
    \end{minipage}
    \hspace{0.04\textwidth}
    \begin{minipage}[b]{0.4\textwidth}
        \centering
        \includegraphics[width=\textwidth]{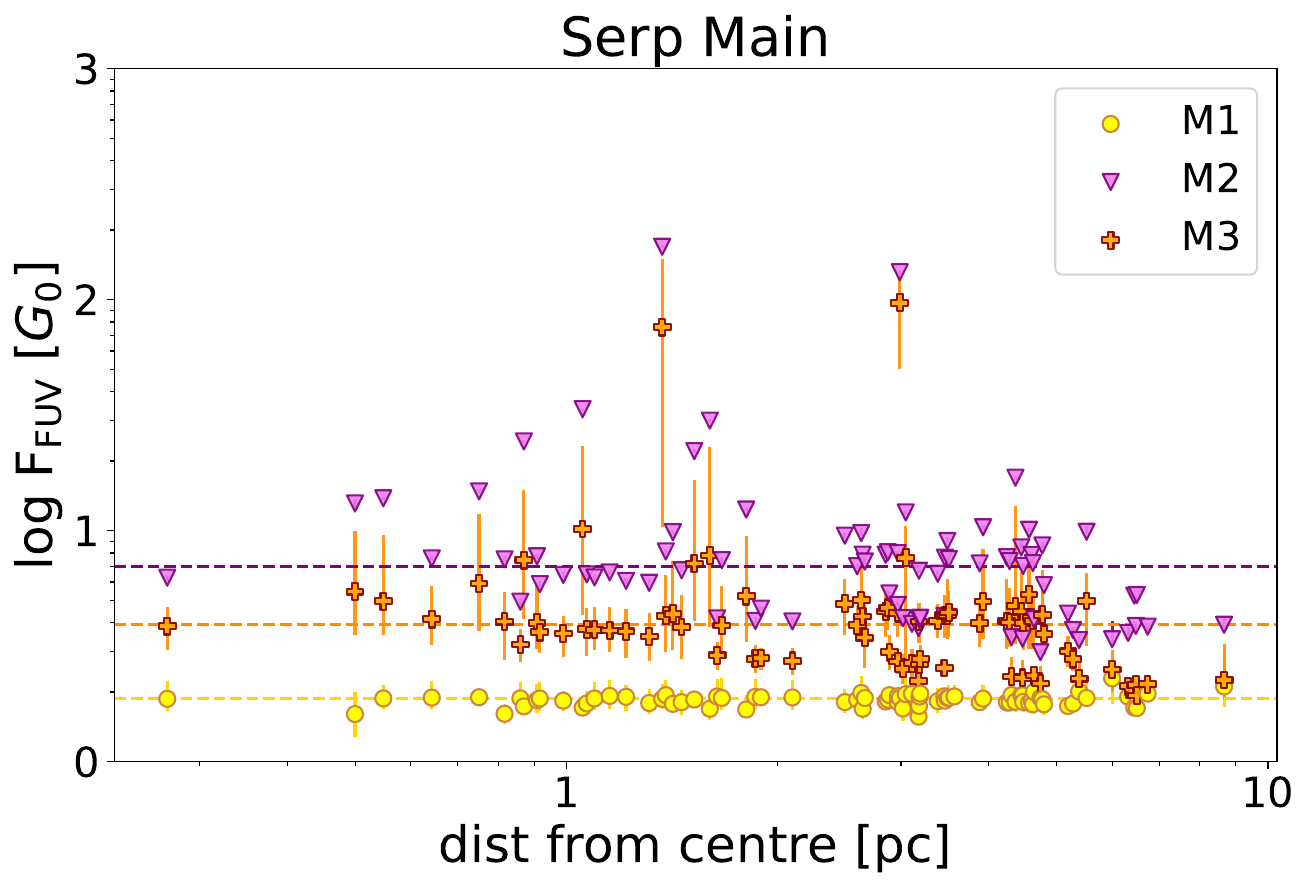}
    \end{minipage}
    \begin{minipage}[b]{0.4\textwidth}
        \centering
        \includegraphics[width=\textwidth]{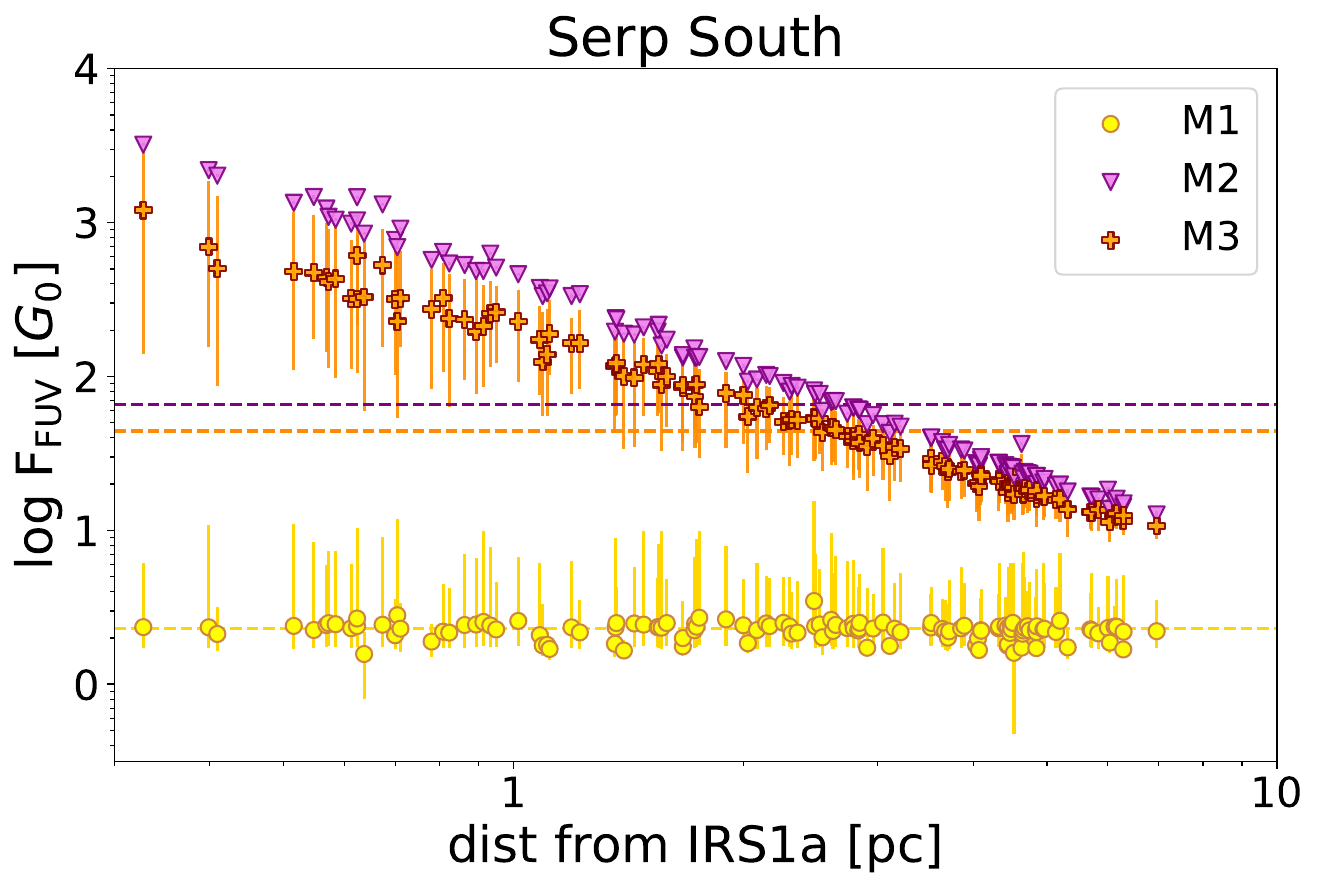}
    \end{minipage}
    \hspace{0.04\textwidth}
    \begin{minipage}[b]{0.4\textwidth}
        \centering
        \includegraphics[width=\textwidth]{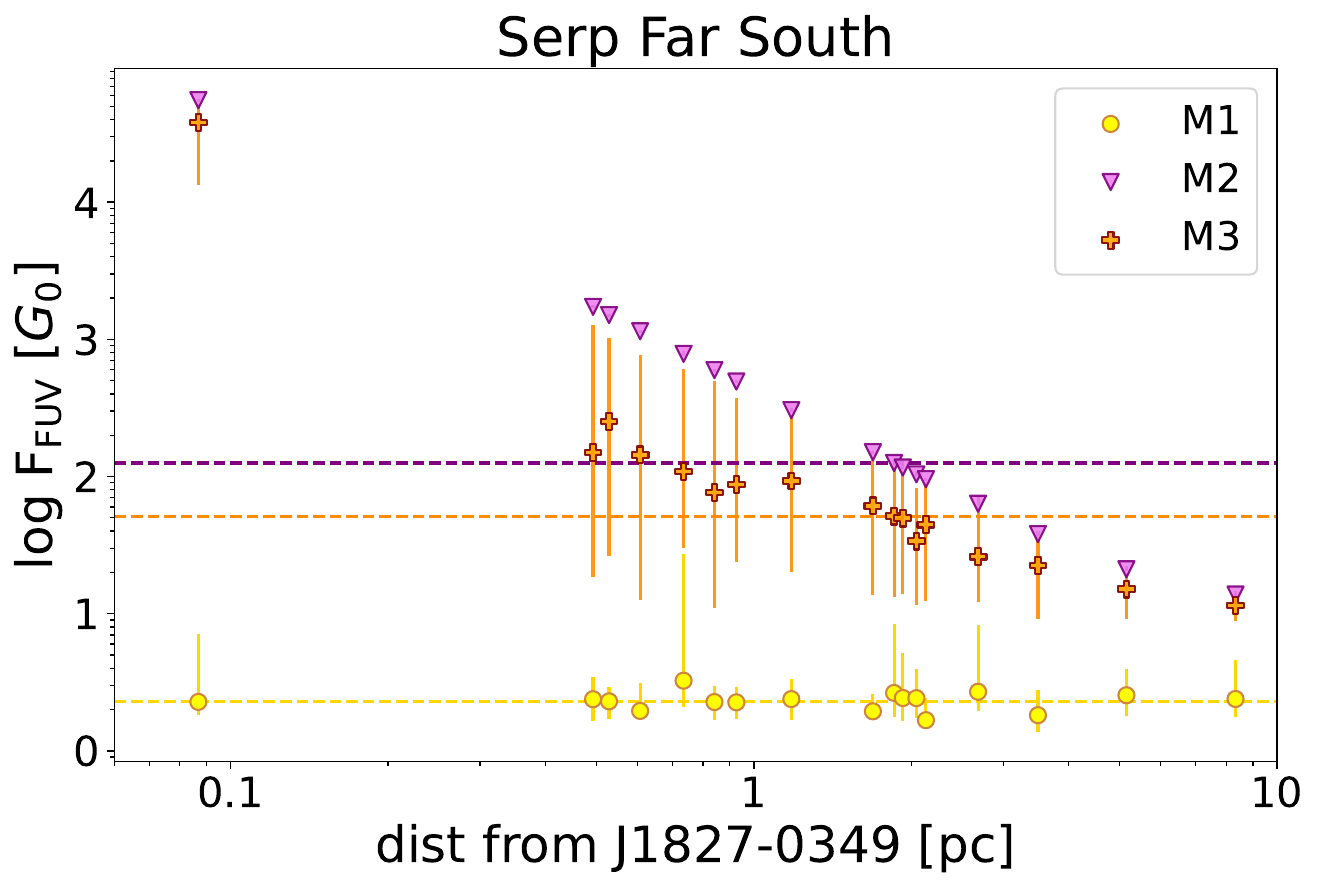}
    \end{minipage}
    \caption{FUV flux experienced by discs in the Serpens region. Based on the sub-region in Serpens where they are located, discs are divided in North East (NE), Main, South, and the most southern part Far South. Our flux estimate using Method 1, Method 2, and Method 3 (Sec. \ref{sec:fuv_calculation_methods_dist_eval}) are shown in yellow, purple, and orange, respectively. Lines show the median fluxes resulting from the three approaches. x-axis: distance from the most massive star, or the centre of the region (if no strong dependence on the most massive star was found).}
    \label{fig:serpens_fuv}
    \end{figure*}
In this Section we present the FUV flux estimate resulting from employing the three approaches introduced in the previous Section, at the position of the star-hosting discs in the selected star-forming regions. We refer to Sec. \ref{subsec:200_pc_regions} and \ref{subsec:orion_serpens_regions} for investigation on the most nearby regions, and regions in Orion and Serpens, respectively.
The extended list of FUV fluxes and uncertainties for the considered discs in the various star-forming regions is available in Table \ref{tab:big_table_of_fluxes}, in Appendix \ref{appendix:big_table}.
In Section \ref{subsec:dust_masses} we investigate the correlation between FUV flux and disc dust mass.

\subsection{Regions within 200 pc from the Sun}\label{subsec:200_pc_regions}
The nearest star-forming regions included in this work contain few late B-type and early A-type stars. Therefore, we expect the FUV irradiation to be lower than in more distant and massive regions.
Moreover, due to their proximity to Earth, the line-of-sight distance of these stellar objects is known with greater precision, which should result in negligible differences when applying the three different methods for the flux calculation.
However, if one or more of the few massive stars are in the close proximity of discs, the resulting FUV irradiation may not be negligible and could significantly influence disc evolution.

In Fig. \ref{fig:nearby_fuvs} we show the FUV flux resulting from applying the three approaches described in Sec. \ref{sec:fuv_calculation_methods_dist_eval}, for the discs in the most nearby regions. 
The three methods proposed for the calculation of the FUV flux are providing a conservative estimate (Method 1), overestimate (Method 2), and our best estimate (Method 3).
In order to apply Method 3 (Sec. \ref{subsec:2dstat_3d_method}), we derived the local density distribution function associated with each region by using the membership catalogues from \citet{Luhman_scocen} in the case of Upper Sco and Lupus, \citet{Luhman_lupus} for Lupus, \citet{Luhman_Tau} for Taurus, \citet{ChamI_ChamII_census} for Cham I and Cham II, \citet{rhoOph_census} for $\rho$ Oph, and \citet{CrA_census} for CrA.
The analytic expression that better describes the local density profile of these regions is a double power law truncated close to the length scale of the star-forming region ($\sim$ the width of the region on the sky plane). The parameters of the expression used for each region are listed and discussed in Appendix \ref{Appendix:density_profiles}.\\

The difference between the three methods is evident in Upper Sco, $\rho$ Oph, Cham I and CrA. Upper Sco presents the highest number of B-type stars among the considered regions, $\sim$50 \citep{Luhman_scocen}, and where external irradiation has been demonstrated to substantially contribute in shaping disc properties \citep{AGEPRO_VIII_ext_photoevap}. The massive and hot stars in this region are mostly B8V and B9V, with the sole O9 being $\zeta$ Oph, which is located at a large separation from the considered discs, and therefore is not influencing significantly the total flux. The median FUV flux in Upper Sco provided by our best estimate is $\sim$$20 \ \mathrm{G_{0}}$.
A similar median is found in $\rho$ Oph, which is located close to the Upper Sco region and contains younger protoplanetary discs with large parallax uncertainties (or no parallax measurements) due to the presence of dense material part of the natal molecular cloud. In this region, $\sim90 \%$ of the total irradiation at the position of the star-hosting discs is produced by the multiple system of massive stars $\rho$ Oph itself (B2IV+B2V, \citeauthor{rho_oph_spt} \citeyear{rho_oph_spt}).
At lower irradiation levels, the presence of massive stars in the close proximity of discs is relevant in Cham I and CrA, where the disc irradiation presents a clear dependence on the separation from the closest massive stars, which are HD 97300 (B9V, \citeauthor{HD_chami} \citeyear{HD_chami}) and HD 96675 (B5V, \citeauthor{HD_chami_} \citeyear{HD_chami_}) in ChamI, and HD 176386 (B9V, \citeauthor{HD_cra} \citeyear{HD_cra}) in Cra. These stars are responsible for the high flux registered for a large sample of discs, with respect to the median, as shown in Fig. \ref{fig:nearby_fuvs}. In these cases, it is evident that the distance uncertainty sampling approach tends to underestimate the flux, especially when discs are close to massive stars. This occurs even in clusters with few massive stars and where the line-of-sight distances are known with higher precision compared to more distant regions. 
The star-hosting disc with the highest FUV flux in CrA is 2MASS J19013912-3653292 (HD 176386B), which form a binary system with the B9 star HD 176386. The close proximity to this massive star results in the high flux observed.

Conversely, in Cham II, no significant B-type stars are found near the considered discs, and therefore the overall flux is determined by field stars, non-members of the region, and the best estimate coincides with the result provided by the sole distance uncertainty sampling approach. Taurus and Lupus show similar results, with few discs being very close to B9 stars. 
The complete list of FUV fluxes with associated uncertainties for individual discs, can be found in Table \ref{tab:big_table_of_fluxes} in Appendix \ref{appendix:big_table}.

\subsection{Orion and Serpens}\label{subsec:orion_serpens_regions}
In Fig. \ref{fig:orion_fuvs} and Fig. \ref{fig:serpens_fuv} we present the FUV flux evaluated at the position of the star-hosting discs in the selected star-forming regions in Orion and Serpens.
The flux estimate provided by using our novel approach ( Method 3, Sec. \ref{subsec:2dstat_3d_method}), shown in orange in Fig. \ref{fig:orion_fuvs} and Fig. \ref{fig:serpens_fuv}, is evaluated considering the local density distribution function derived by using known stellar membership of each star-forming region from various catalogues retrieved from literature. Specifically, we used \citet{APOGEE_survey} for the ONC, $\sigma$ Ori, $\lambda$ Ori, and 25 Ori, \citet{Da_Rio_2012} and \citet{Trapezium_census} for the ONC,  \citet{SigmaOri_3D_2024} for $\sigma$ Ori and NGC 2024, \citet{Suarez_25ori_census} for 25 Ori, and the updated sample of members of NGC 1977 by Kim et al. (in prep.). 
In regions presenting a 2D nearly centrally concentrated geometry, which in our sample corresponds to the ONC, $\sigma$ Ori, $\lambda$ Ori, and 25 Ori regions, the local density profile was defined as function of the separation from the centre of the region, as discussed in Sec. \ref{sec:synthetic_clusters}. The analytic expression that better describes this profile is an exponentially truncated single power law, which was already known in previous works for the ONC (e.g. \citeauthor{Da_Rio_2012} \citeyear{Da_Rio_2012}). 
In the case of NGC 2024, the confirmed stellar census contains too few stars complementing the star-hosting discs included in our study, spread across various separations, making it difficult to recover enough stars per unit separation to derive a reliable density profile. Therefore, our statistical approach cannot be fully trusted, and we present only the FUV fluxes obtained through Method 1 (Sec.\ref{subsec:3d_method}), and Method 2 (Sec. \ref{subsec:2d+3d_method}). 
Additional details on the derivation and reliability of the density profile, along with the specific parameters used in the calculation, are discussed in Appendix \ref{Appendix:density_profiles}.\\

For the Serpens region we used the Gaia DR2 membership census by \citet{Serpens_gaiadr2} and the Young Stellar Objects (YSOs) catalogue of \citet{Anderson22_Serpens} to derive the local density profile of the three main extended sub-regions, which are North East (NE), Main, and South.
In Fig. \ref{fig:serpens_fuv}, the investigated ClassII discs in the most southern region of Serpens South, referred to as Far South, are showed separately from the others. This distinction highlights the dependence of the flux on the separation from the most massive stars, which are IRS1a (O9.5V) and 2MASS J18273952-0349520 (B1.5V, \citeauthor{serpens_massive} \citeyear{serpens_massive}), respectively located in Serpens South (near the HII region Westerhout 40, \citeauthor{W40_serpens} \citeyear{W40_serpens}), and Far South. The highest level of irradiation is found near the HII region Westerhout 40 (W40), where a small group of massive stars is concentrated (\citeauthor{Anderson22_Serpens} \citeyear{Anderson22_Serpens}).
Refer to Appendix \ref{Appendix:density_profiles} for the 2D sky maps of the selected disc sample and neighbouring massive stars included in this work, corresponding to each star-forming region.

The average FUV flux is significantly higher in the considered Orion and Serpens regions (with exception of Serp NE and Serp Main) compared to the closest regions (Fig. \ref{fig:serpens_fuv}). The flux estimate provided by involving Method 3 reaches $\sim10^{5} - 10^{6} \ \mathrm{G}_{0}$ in the ONC, where the total flux is dominated by the most luminous O-type and early B-type stars, while field stars contribute on average only about $\sim1 \%$ of the total flux.
Method 1 (Sec. \ref{subsec:3d_method}) results in a nearly flat distribution of FUV fluxes in all the regions included in this study, with median value $\lesssim 10 \ \mathrm{G_{0}}$ (as showed in yellow in Fig. \ref{fig:nearby_fuvs}, \ref{fig:orion_fuvs}, and \ref{fig:serpens_fuv}).

Both using Method 2 and Method 3 demonstrate a clear dependence of the flux on the distance from the most massive and luminous star in the region, as shown on the x-axis of the panels in Fig. \ref{fig:orion_fuvs}. 

As for the synthetic cluster, we show that Method 1 using real parallax errors from Gaia DR3 and Hipparcos severely underestimates the FUV flux, while placing constraints based on the 2D geometry of a star-forming region can result in a more accurate FUV estimate.

\subsection{Correlation with Disc Properties: Dust Disc Mass}\label{subsec:dust_masses}
Externally irradiated discs can be significantly depleted in both gas and dust component (e.g. \citeauthor{Owen_2012} \citeyear{Owen_2012}, \citeauthor{Sellek_dust_mass_2020} \citeauthor{Sellek_dust_mass_2020}). In order to understand the impact of external photoevaporation on disc evolution and composition, it is crucial to investigate the relation between disc properties and the external radiation they experience.
In this Section we focused on the dust disc mass, as gas disc mass and disc size are more challenging to determine and available only for few discs in regions at distances larger than $\sim$200 pc.
\begin{figure}[htbp]
    \centering
    \includegraphics[width=0.4\textwidth]{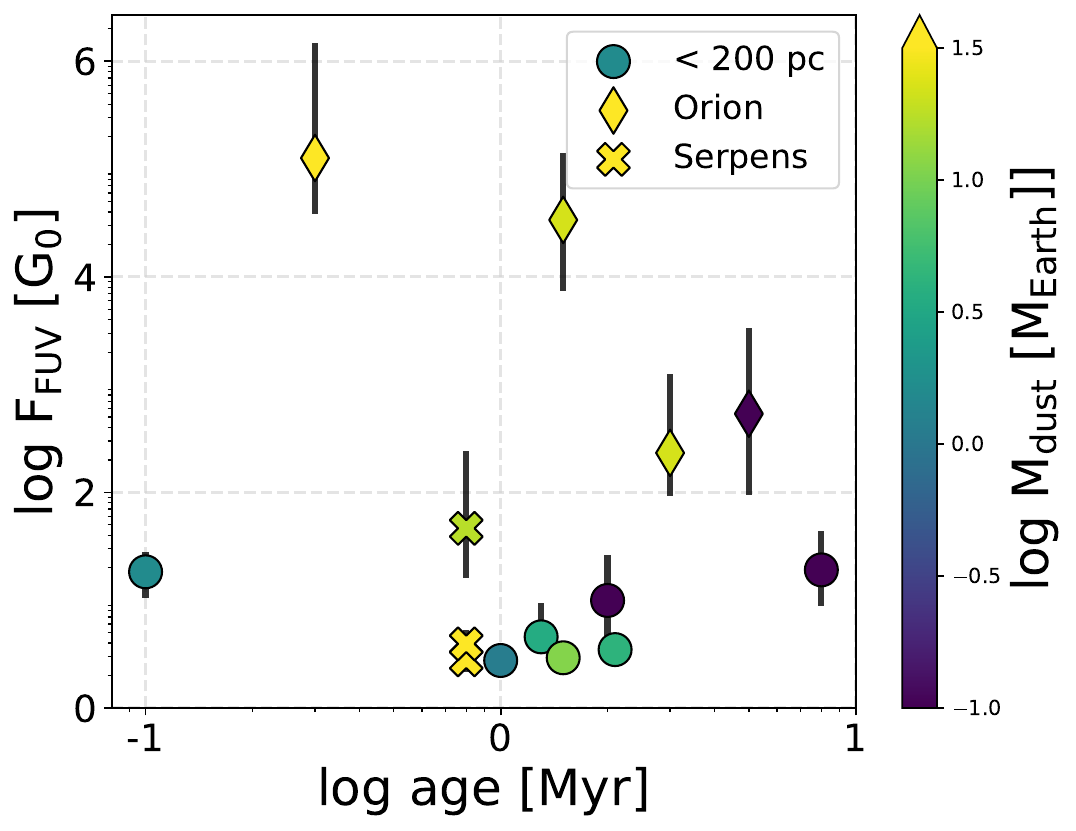}
    \hspace{0.04\textwidth}
    \caption{Median FUV flux in function of the estimated age of the regions. Each dot represents a star-forming region in this work, and is colour-coded based on the median dust disc mass available for the sample of discs considered. NGC 1977 and 25 Ori are not included, due to lack of dust disc mass measurements. The FUV flux presented for NGC 2024 is evaluated using Method 2 (Sec. \ref{subsec:2d+3d_method}).}
    \label{fig:age_vs_fuv_vs_mdust}
\end{figure}
\begin{figure}[ht]
    \centering
    
    \begin{minipage}[b]{0.35\textwidth}
        \centering
        \includegraphics[width=\textwidth]{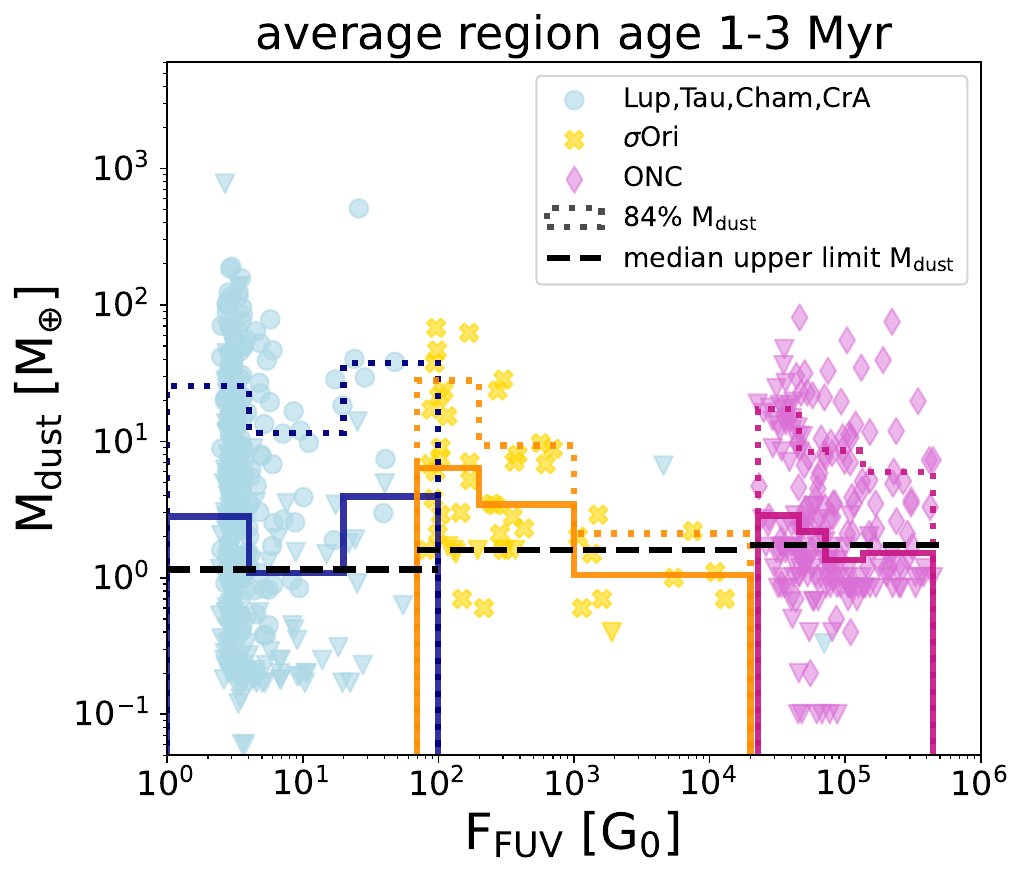}
    \end{minipage}
    \begin{minipage}[b]{0.35\textwidth}
        \centering
        \includegraphics[width=\textwidth]{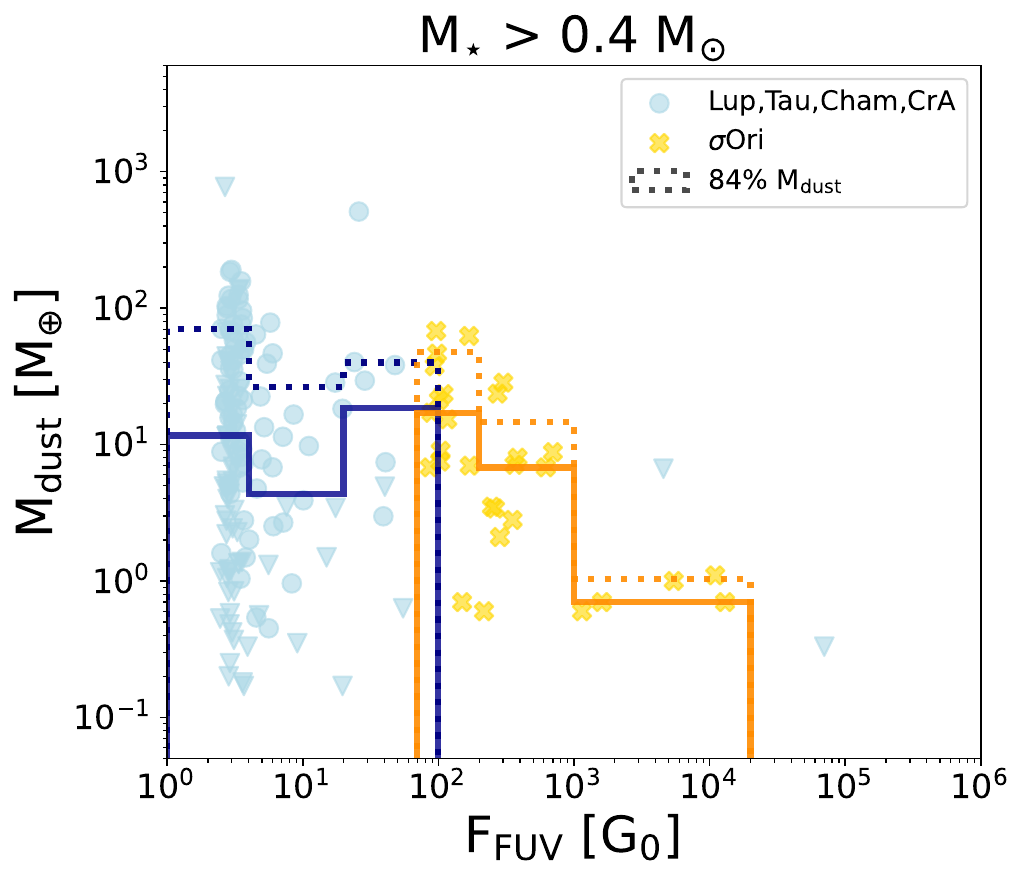}
        \vspace{0.01cm}
    \end{minipage}
    \begin{minipage}[b]{0.35\textwidth}
        \centering
        \includegraphics[width=\textwidth]{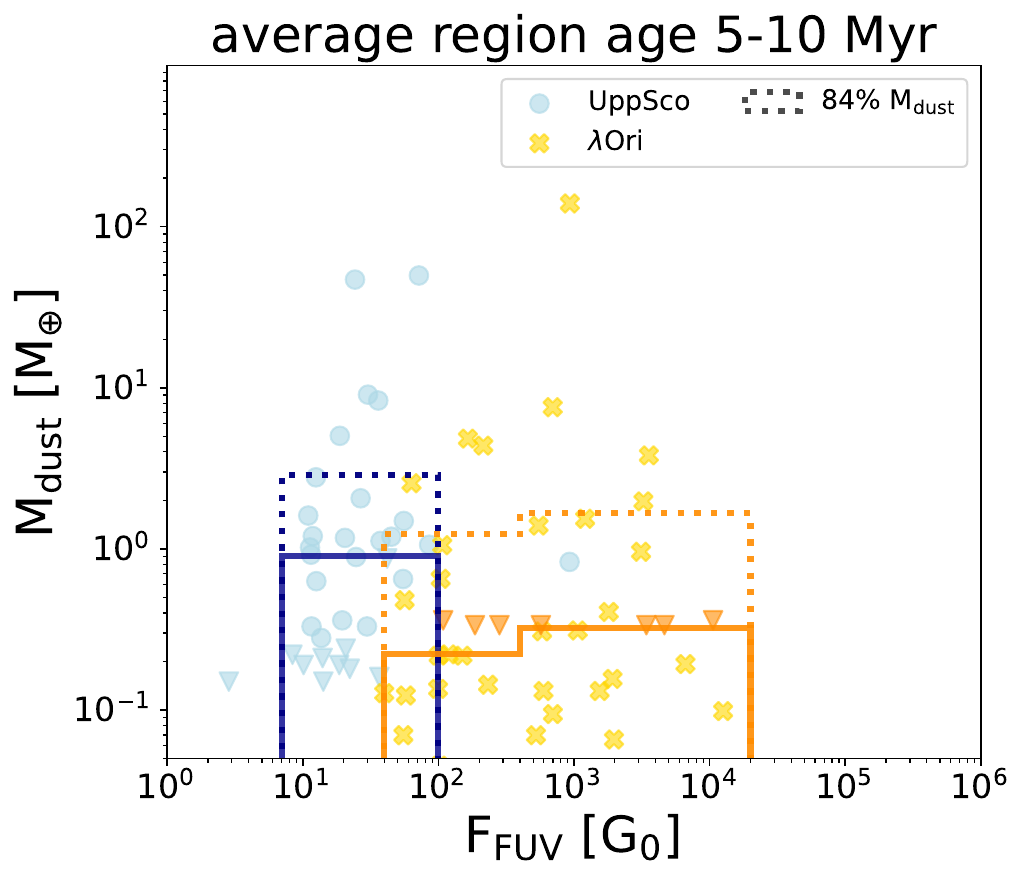}
    \end{minipage}
    \caption{Dust disc masses and FUV fluxes for the regions with average age 1-3 Myr (top and central panel), and 5-10 Myr (bottom panel). Histograms show medians dust disc masses in flux bins, including upper limits (shown in coloured triangles). Dotted histograms contains the 84\% of the distributions. Top panel presents the median upper limit dust mass in each region as black dashed lines. Central panel includes only the objects with stellar mass > 0.4 M$_{\odot}$, and no upper limit in dust disc mass is present in $\sigma$ Ori. The ONC is not included in this plot as the stellar mass is defined only for $\sim$1/3 of the total disc sample. }
    \label{fig:mdust_fuv_histo}
\end{figure}

Dust content in discs is a more accessible quantity and can be associated to the thermal emission at (sub-)mm wavelengths, assuming optically thin emission and setting a certain dust temperature and opacity (e.g. \citeauthor{Miotello_PPVII} \citeyear{Miotello_PPVII}). 
Dust disc mass as retrieved from continuum observations is available for part of the investigated disc sample in 12 over 14 regions included in this work (25 Ori and NGC 1977 discs currently lack dust mass measurements). 
We selected from each of these 12 regions the discs with measured dust mass.
In Fig. \ref{fig:age_vs_fuv_vs_mdust} we show the median FUV flux as a function of an average age of the region, colour-coded based on the median dust disc mass. 
The uncertainties on the FUV flux are 16$^{\mathrm{th}}$ and 84$^{\mathrm{th}}$ percentiles of the distributions. 
The median FUV flux of NGC 2024 is obtained by using 2D projected separations from the OBA stars members of the region,
as the local density profile of the region, needed to apply the method proposed in this study, cannot be reliably derived here. However, since the NGC 2024 disc sample presents dust mass measurements, we decided to show the results taking into account that the median flux value is an overestimate of the actual flux.

Figure \ref{fig:age_vs_fuv_vs_mdust} shows that less massive discs are generally observed in older regions, a trend that has been extensively investigated by disc surveys in the most nearby star-forming regions (e.g. \citeauthor{Manara_2023} \citeyear{Manara_2023}). However, protoplanetary discs in more and less irradiated regions are expected to evolve differently, and therefore disc properties across different regions cannot be compared on the sole basis of the age, but the irradiation from the environment should also be considered.
Theoretical and numerical models predict that discs exposed to stronger external irradiation deplete more quickly (e.g. \citeauthor{winter_photoev} \citeyear{winter_photoev}), and therefore we would expect to observe less massive discs in higher irradiated environments.
Observations within individual star-forming regions, such as $\sigma$ Ori and the ONC, support this prediction, showing a clear dependence of disc mm flux emission on FUV flux, where the FUV flux values involved are obtained considering 2D projected separations from the closest massive stars (\citeauthor{Ansdell_2017_sori} \citeyear{Ansdell_2017_sori}, \citeauthor{Mauco_2023_sori} \citeyear{Mauco_2023_sori}, \citeauthor{Eisner_2018_onc} \citeyear{Eisner_2018_onc}). This raises the question of whether a similar correlation might hold on a larger scale, across multiple star-forming regions, using more accurate estimate of the FUV fluxes.
Firstly, we investigated this potential correlation on a region-averaged perspective, considering median dust disc mass and the FUV fluxes (evaluated using Method 3, Sec. \ref{subsec:2dstat_3d_method}) in each region, as presented in Fig. \ref{fig:age_vs_fuv_vs_mdust}. Subsequently, we focus our study on the distribution of dust disc masses and FUV fluxes across regions of similar age, as showed in Fig. \ref{fig:mdust_fuv_histo}.

No clear correlation between median dust mass and median FUV flux was identified (Fig. \ref{fig:age_vs_fuv_vs_mdust}). However, several considerations need to be addressed. First of all, the regions in Orion and Serpens presenting high median FUV fluxes in Fig. \ref{fig:age_vs_fuv_vs_mdust}, very young. This could explain the high dust mass observed with respect to what measured in the most nearby regions, as external photoevaporation did not have enough time to significantly remove disc material. Moreover, interstellar dust can efficiently shield discs from high-energy external UV photons during disc first half million year \citep{Ali_2019}, and therefore the detected discs could have experienced intense UV radiation only very recently.
However, in more distant regions observations are biased toward detecting more massive discs, leading to median dust disc masses which are not entirely representative of the regions. 
Additionally, the stellar mass distribution is also not homogenous across Serpens, Orion and the nearest regions, with more distant regions presenting only few objects identified as low-mass stars ($\lesssim$0.3 M$_{\odot}$) and many more missing stellar mass measurements. 

Figure \ref{fig:age_vs_fuv_vs_mdust} shows a clear lack of information on dust disc mass at intermediate-to-high irradiation levels (> $10^{2} \ \mathrm{G}_{0}$) with respect to less irradiated regions. \\

So far we considered region-averaged values, now we focus on individual values, dividing the disc sample in age bins, considering regions with similar average age as discs disperse over time. Specifically, we considered the age range $\sim$1-3 Myr, where more than two regions present available dust masses. The selected regions are Lupus, Taurus, Cham I, Cham II, CrA, $\sigma$ Ori, and the ONC. 
Due to the high concentration of dust mass measurements at low FUV fluxes compared to intermediate and high irradiation levels, we divided our dust disc mass sample into flux bins of varying width, showing the distribution of the median dust masses (in each bin) against the distribution of fluxes, in the histograms in Fig. \ref{fig:mdust_fuv_histo}. Bin widths are chosen based on the number of objects and upper limit dust masses in each region. Specifically, the sample of the most nearby discs, covering $\sim$1-100 $\mathrm{G}_{0}$, is divided into three parts, as well as the $\sigma$ Ori sample, which covers fluxes from $\sim$$10^{2}$ to $\sim$$2\times 10^{4} \ \mathrm{G}_{0}$. The ONC disc sample, which consists of detections and non-detections in \citet{Eisner_2018_onc}, is divided into $\sim$40 stars each, with at least four detections in each bin. Since in the ONC $\sim$60\% of the total disc sample is undetected, the median dust disc mass is dominated by upper limit dust masses (shown in black dashed lines in the top panel of Fig. \ref{fig:mdust_fuv_histo}) and is comparable to the observed median in the other regions.
Due to the known dependence of the dust mass on the stellar mass (e.g. \citeauthor{Manara_2023} \citeyear{Manara_2023}), the medians can be highly influenced by the range of stellar masses covered.
In order to reduce the effect produced by the dependence on stellar mass, we restricted the disc sample in these regions to the relatively high-mass stars ($M_{\star} > 0.4 \ \mathrm{M}_{\odot}$), which are the majority detected in $\sigma$ Ori. The results are shown in the central panel of Fig. \ref{fig:mdust_fuv_histo}. We did not include the ONC in this panel as stellar mass is defined only for $\sim$1/3 of the total disc sample.
A decrease in dust disc mass is visible between $\sim$$10^{3}$ and $\sim$$10^{4} \ \mathrm{G}_{0}$, and is related to $\sigma$ Ori (Fig. \ref{fig:mdust_fuv_histo} central panel), where this trend has been previously found  (\citeauthor{Ansdell_2017_sori} \citeyear{Ansdell_2017_sori}, \citeauthor{Mauco_2023_sori} \citeyear{Mauco_2023_sori}), and now supported by the use of our updated FUV flux estimate. The correlation in $\sigma$ Ori is more pronounced when restricting to high stellar masses (Fig. \ref{fig:mdust_fuv_histo}). 
A weaker negative dependence of the dust mass on the external flux appears in the ONC
, more clearly visible when looking at the 84$^{\mathrm{th}}$ percentile of the mass distribution, as the median is highly influenced by the undetected objects. 
The lack of stellar properties and disc mm continuum detections at intermediate-to-high irradiation levels limits our ability to draw more exhaustive conclusions across a wide range of distances, regions, and FUV fluxes. Therefore, we strongly support explorations of this parameter space in future observations.
In Appendix \ref{appendix:corr_mdust_fuv}, Fig. \ref{fig:corr_fuv_mdust_mstar}, we show that we found a tentative correlation between individual dust disc masses and FUV fluxes for objects in the range of ages 1-3 Myr. However, the lack of information on stellar mass and dust disc mass above $10^{2} \ \mathrm{G}_{0}$ prevents us from strongly support the correlation derived.

When applying the same study to the oldest regions of our sample, which are Upper Sco and $\lambda$ Ori, we only detect a small decrease in median dust mass from lower to higher FUV fluxes (from Upper Sco to $\lambda$ Ori). The results are shown in the bottom panel of Fig. \ref{fig:mdust_fuv_histo}.  
As the disc sample is limited to only two regions, there is insufficient data for a deeper statistical investigation to disentangle if the small decrease in dust disc mass is due to external FUV irradiation or observational biases. Future dust mass measurements of 25 Ori discs will offer crucial data to further explore the relation.

\section{Discussion}\label{sec:discussion}

\begin{figure*}[htbp]
    \centering
    \includegraphics[width=\textwidth]{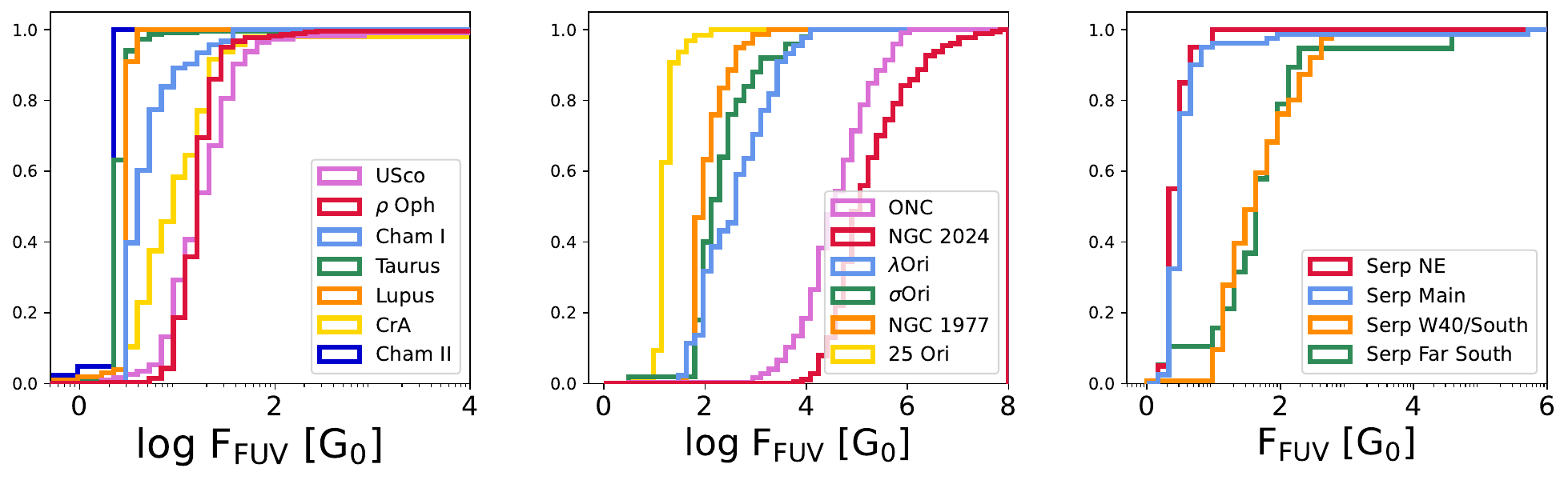}    
    \caption{Cumulative distributions of the median FUV fluxes experienced by discs in the regions selected in $\lesssim200$ pc (left panel), Orion (central panel), and Serpens region (right panel). The flux values used are the best estimate evaluated using Method 3 (Sec. \ref{subsec:2dstat_3d_method}). The exception is NGC 2024, where we evaluated FUV fluxes using Method 2 (Sec. \ref{subsec:2d+3d_method}), considering them as 
    overestimate of the actual fluxes.}
    \label{fig:cumulative_distr}
    \end{figure*}
\begin{figure}[htbp]
    \centering
    \includegraphics[width=0.5\textwidth]{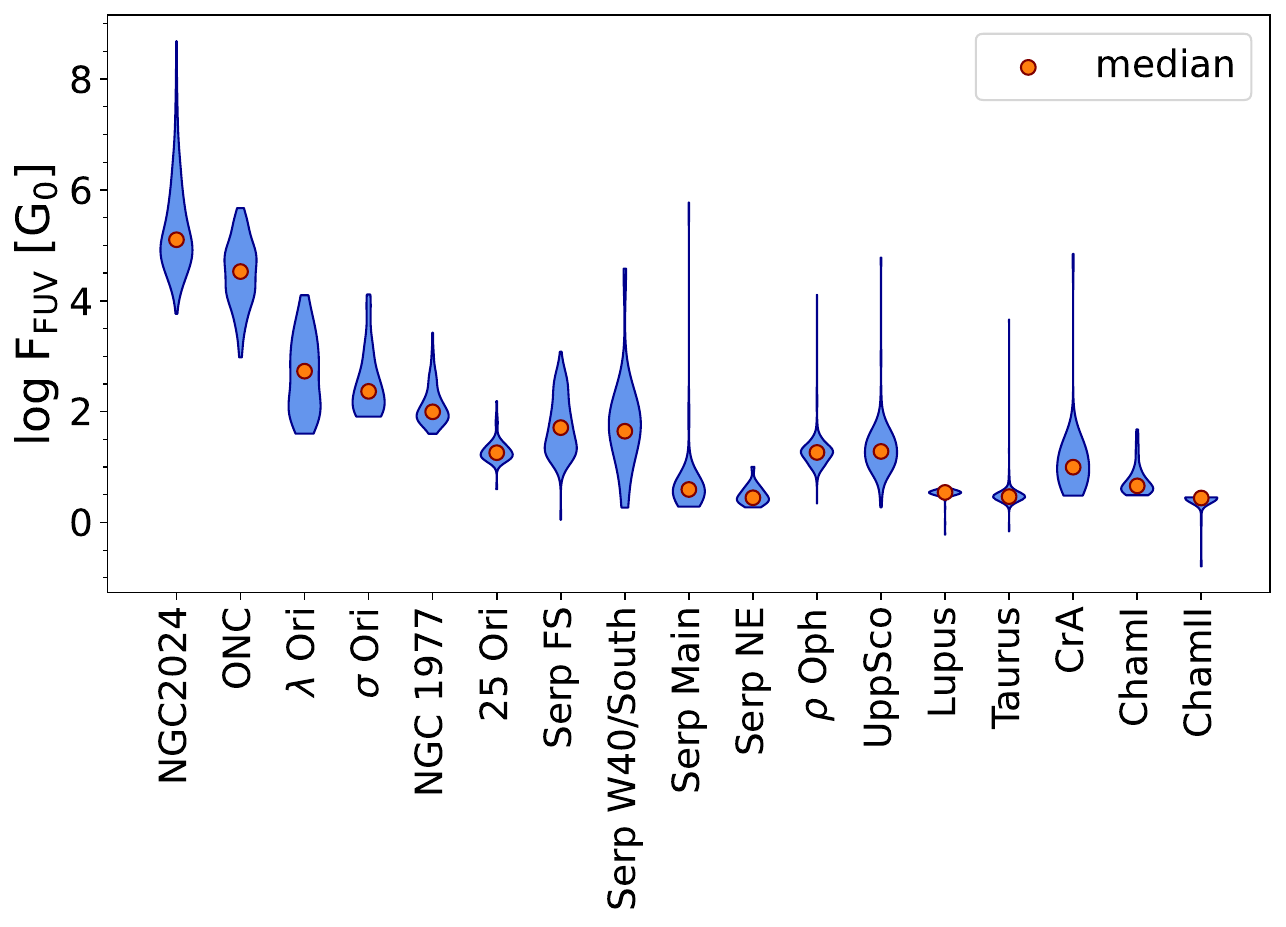}
    \hspace{0.04\textwidth}
    \caption{Distributions of the FUV fluxes for the star-forming regions included in this work. The median of the distributions are shown in orange. NGC 2024 fluxes are overestimate of the actual values, as we used Method 2 (Sec. \ref{subsec:2d+3d_method}) to compute the FUV flux, due to the lack of information to accurately derive the local density profile.}
    \label{fig:violin_plot_tot}
\end{figure} 
In this study we used the available information on the 2D geometry of a star-forming region to compute the best estimate of the FUV flux at the position of a sample of star-hosting discs (Method 3, Sec. \ref{subsec:2dstat_3d_method}). 
In this Section, we discuss the FUV fluxes obtained for each region (Sec. \ref{subsec:discussion_results}), the uncertainties, limitations, and future improvements of our proposed method (Sec. \ref{subsec:limitations}), and the correlation with disc properties (Sec. \ref{subsec:discuss_disc_properties}). 

\subsection{FUV Flux in the Star-Forming Regions Considered}\label{subsec:discussion_results}
Figure \ref{fig:cumulative_distr} presents the cumulative distributions of the resulting FUV fluxes in the various regions, respectively divided in the most nearby regions (left panel), regions in Orion (central panel), and Serpens region (right panel), for a better visualisation. In Fig. \ref{fig:violin_plot_tot}, we provided, for each region included in this study, the distribution of median FUV fluxes at the position of star-hosting discs. 
Discs in Orion experience the highest levels of irradiation among the studied regions, particularly those in the ONC and NGC 2024\footnote{We recall that for NGC 2024 we used the flux values derived from using Method 2 (Sec. \ref{subsec:2d+3d_method}), because the local density profile cannot be reliably derived in this region. These values are overestimate of the actual FUV flux.}. However, we verified that, even in regions where the median irradiation is a hundred times lower than in the ONC, the flux distributions still reach moderate and high levels ($10^{2} - 10^{3} \ \mathrm{G}_{0}$). This is due to late B-type and early A-type stars closely located to star-hosting discs, and underlines the importance of considering the external irradiation even in moderately irradiated regions.
The median incidental external FUV flux on disc-hosting stars in the lowest irradiated regions, which are ChamII, Taurus, Serp NE, and Serp Main, is $\sim$2-3$ \ \mathrm{G}_{0}$, which is induced by field massive stars, not members of the star-forming regions themselves. Upper Sco and $\rho$ Oph present in median the highest FUV irradiation field across the most nearby regions, due to the presence of numerous massive stars in the Sco-Cen OB association \citep{OB_associations_wright}.
Disc truncation due to external photoevaporation can efficiently reduce disc mass and size (with consequences on planet formation and composition), even below 100 $\mathrm{G}_{0}$, as demonstrated by numerical models and the comparison with observations \citep{AGEPRO_VIII_ext_photoevap}. Therefore, being aware of the irradiation level in the most nearby regions is essential to interpret observations. 
The median FUV fluxes on star-hosting discs in star-forming regions in Orion range from $\sim20 \ \mathrm{G}_{0}$, which is registered in 25 Ori, to $\sim5\times10^{5} \mathrm{G}_{0}$ in the ONC. The FUV flux in Serpens is dominated by the irradiation produced by the O-type star and the early B-type stars close to Serp South and the HII region W40.

\begin{table}[h]
    \caption{\label{tab:fuv_medians}Median FUV flux for the regions included in this work}
    \centering
    \vspace{0.2cm}
    \def\arraystretch{1.6}
    \begin{tabular}[H]{|P{1.8cm}| P{2.5cm}| P{2.5cm}|}
    \hline
    \hline
    Region    & F$_{\mathrm{FUV, median}}$ & $( \sigma_{\mathrm{FUV}} / \mathrm{F}_{\mathrm{FUV}} )_{\mathrm{median}}$ \\
     & $[\mathrm{G}_{0}]$ & \\
    \hline 
    UppSco     & 20.11$^{+21.03}_{-10.7}$  & 0.35 \\
    Lupus      & 3.48$^{+21.03}_{-0.31}$     &  0.06   \\
    $\rho$ Oph    & 18.33$^{+7.63}_{-7.12}$   & 0.30   \\
    Taurus     & 2.92$^{+0.16}_{-0.14}$     & 0.05  \\
    Cham I     & 4.57$^{+4.3}_{-1}$   & 0.19   \\
    Cham II    & 2.76$^{+0.03}_{-0.02}$     & 0.08   \\
    CrA        & 9.94$^{+14.83}_{-5.69}$    & 0.39   \\
    ONC        & 33713.93$^{+97361.13}_{-25803.39}$ & 0.96  \\
    $\sigma$ Ori  & 232.54$^{+946.43}_{-103.7}$  & 0.48 \\
    $\lambda$ Ori & 536.86$^{+2602.54}_{-436.50}$   & 0.57 \\
    NGC 2024 & 126462.61 &  - \\
    NGC 1977   & 100.11$^{+195.16}_{-32.07}$ & 0.73   \\
    25 Ori     & 18.14$^{+8}_{-3.86}$    & 0.26   \\
    Serp Main  & 3.93$^{+1.02}_{-1.42}$    & 0.22   \\
    Serp NE    & 2.79$^{+1.46}_{-0.49}$   & 0.13   \\
    Serp South & 44.27$^{+190.37}_{-26.61}$   & 0.37  \\
    \hline
    \end{tabular}
    \tablefoot{For each region listed in Column 1, we indicate median FUV flux and uncertainties (16$^{\mathrm{th}}$ and 84$^{\mathrm{th}}$ percentiles of the distributions of median individual FUV fluxes), and median relative uncertainty. The FUV flux is evaluated using Method 3 (Sec. \ref{subsec:2dstat_3d_method}), with exception being NGC 2024, for which we indicate the upper limit flux evaluated using Method 2 (Sec. \ref{subsec:2d+3d_method}), and Cham II, which does not contain any relevant OBA star and therefore the FUV flux estimate refers to Method 1 (Sec. \ref{subsec:3d_method}).}  
\end{table}
\subsection{Uncertainties, Limitations of the Model, and Future Improvement}\label{subsec:limitations}
The primary sources of uncertainty in the flux calculation are the relative (3D) distances between massive stars and star-hosting discs, and the FUV luminosity of massive stars.
In our study we neglected in most cases the latter, and we now discuss why this is justified, and show that this causes a small effect.

In the flux calculation, the uncertainty in FUV luminosity is caused by the uncertainty in the effective temperature of massive stars. Since most of the OBA stars included in this work have effective temperature defined in the Astrophysical Parameters (ESPHS) pipeline provided by Gaia DR3, we have access to the temperature uncertainty and we used it to investigate the implications on the final FUV flux value.
In the ESPHS pipeline, the error in effective temperature increases with spectral type, reaching unreliable temperature values and uncertainties above 25000 K \citep{GaiaDR3_ESPHS_2}.
From stellar evolution models we retrieved that effective temperatures between 8000 and 25000 K correspond to stellar masses between $\sim$2 and $\sim$10 M$_{\odot}$ (assuming an age of 1 Myr), and FUV luminosities of $\sim$$10^{34}$ - $10^{37}$ erg s$^{-1}$.
Massive stars below 23000 K have typical errors on effective temperatures of a few tens of Kelvin,
which translates into errors of a few percent in FUV luminosity and FUV flux (at a fixed separation). These are negligible errors compared to the uncertainty associated with relative separations.
Massive stars in the temperature range [23000 - 25000] K reach temperature uncertainties of few thousand Kelvin, which result in a FUV flux variation of $\sim$$80$\%.  
In the last case, where errors cannot be neglected, a more precise temperature estimates is provided by spectral classification tables (\citeauthor{Gray_Corbally_2009} \citeyear{Gray_Corbally_2009}). 
Therefore, in order to reduce the impact of the uncertainty in FUV luminosity on the total flux, in our calculation we retrieved the effective temperature of massive stars above 23000 K (which might present highly uncertain Gaia ESPHS effective temperature) from spectral classification, as introduced in Sec. \ref{sec:fuv_calculation:sample_star}. \\

In this study we proposed the use of the local density profile of a star-forming region to deal with uncertainties in parallax measurements and determine the 3D separation between pairs of stars in a star-forming region. However, an intrinsic limitation of Method 3 (Sec. \ref{subsec:2dstat_3d_method}) comes from the shape of the local density function itself, which is individually derived (and analytically prescribed) for each region. 
Specifically, the density profile was used into the probability function in Eq. \ref{eq:prob_rR_final}, from which we sampled the Cumulative Distribution Function (CDF) to determine the best estimate of the 3D separation between stars. Therefore, an extremely flat density profile results in a nearly uniform probability of 3D separations across a wide range of values ($\sim$flat CDF). As a consequence, the uncertainty on the final flux will be large. 
Conversely, if the initial density profile is extremely steep, the corresponding inverse CDF will be steep as well, presenting its maximum close to the 2D separation value, leading to a high probability of obtaining fluxes similar to those predicted using the pure 2D separation.
Table \ref{tab:fuv_medians} lists median FUV flux values, with 16$^{\mathrm{th}}$ and 84$^{\mathrm{th}}$ percentiles of the distribution of median FUV fluxes, and median relative uncertainty for each region.
Relative uncertainties in the various regions generally do not depend on $\mathrm{G}_{0}$. However, in the case of centrally symmetric regions, high FUV flux values can present large error bars. This results because a higher $\mathrm{G}_{0}$ corresponds to a smaller separation between disc and massive stars, which corresponds to the $\sim$flat part of the spherically symmetric local density function (see Plummer sphere in Sec. \ref{sec:synthetic_clusters}). 

While errors produced by the shape of the local density distribution cannot be removed, future improvements can be made for a better identification of the density profile of star-forming regions. Deriving accurate local density profiles requires membership census of star-forming regions numerous enough to present a substantial number of stars per unit separation, which permits to resolve stellar over-densities that influence the shape of the local density function.
Therefore, future stellar membership studies in the regions included in this work will improve the precision of the FUV flux calculation method. Moreover, new stellar membership catalogues in additional regions will allow the application of the method described in this study to a larger stellar sample. 

\subsection{FUV Flux and Disc Properties}\label{subsec:discuss_disc_properties}
In Sec. \ref{subsec:dust_masses} we focused on the investigation of the relation between FUV flux estimate at the position of star-hosting discs and dust disc masses, as dust disc size and gas disc mass and size is not measured for the majority of the discs at distances larger than $\sim$200 pc.
Among the regions considered in this work, only $\sigma$ Ori and Upper Sco show a decrease in dust and gas disc size at higher $\mathrm{G}_{0}$, as presented in \citet{Mauco_2023_sori} and \citet{AGEPRO_VIII_ext_photoevap}. 
Figure \ref{fig:mdust_fuv_histo} highlights key areas of the FUV flux-age parameter space to explore with future observations in order to deeply investigate how external irradiation contributes in shaping dust disc content. Specifically, more stellar information (in particular stellar mass measurements) and deeper disc observations in intermediately and highly irradiated environments (> $10^{2} \ \mathrm{G}_{0}$), in a range of ages 1-10 Myr, would be fundamental to complement our study.
Therefore, we strongly encourage measurements of disc properties in other star-forming regions, in order to explore the relation with the environmental irradiation.  

In proplyds, direct information on the disc mass loss rate due to external photoevaporation is provided by ionised gas emission lines (e.g. [NII], [Ne III],  [OIII]), which are bright in PDRs and seen at optical wavelengths.
One of the most sensitive lines to external UV radiation and good tracer of the inner region of the wind is [OI] 6300 $\AA$ (e.g. \citeauthor{Bally_OI}\citeyear{Bally_OI}, \citeauthor{Storzer_OI}\citeyear{Storzer_OI}),  whose luminosity is predicted to increase with external FUV field strength (e.g. \citeauthor{Ballabio_2023} \citeyear{Ballabio_2023}).
However, at low irradiation levels ($\lesssim$$10^{3} \ \mathrm{G}_{0}$) the line luminosity is dominated by the irradiation from the central star, and cannot be used as tracer of external winds. The observed line luminosity is therefore similar across nearby regions and $\sigma$ Ori \citep{Mauco_2024}. More irradiated regions included in this work, such as the ONC, have too few or missing detections for a statistically reliable study on optical lines.
Physical and chemical characteristics of PDRs are also provided by models and observations of infrared (IR) lines (e.g. rovibrational lines of H$_{2}$ and CO, mid-IR features of polycyclic aromatic hydrocarbons \citeauthor{Hollenbsch_H2}\citeyear{Hollenbsch_H2}), and are now investigated in detail by JWST in the solar neighbourhood (e.g. \citeauthor{PDR4all_H2}\citeyear{PDR4all_H2}). However, the problem of using this lines to trace external photoevaporation of intermediately and moderately irradiated protoplanetary discs remains, as the radiation from the central star will obscure the contribution of an external UV source.
This highlights the need for, on one hand, an independent method to evaluate the FUV radiation in PDR models to accurately compare with observations and, on the other hand, alternative PDR tracers in the absence of proplyd-like morphology.

\section{Conclusions}\label{sec:conclusions}
In this work we proposed a novel approach to evaluate the FUV flux, along with its uncertainty, at the position of star-hosting discs in nearby star-forming regions. We analysed three methods to deal with the uncertainty in separation between massive stars and low-mass stars in 3D space, which is the main source of uncertainty in flux calculations. These methods consist of: distance uncertainty sampling (Method 1, Sec. \ref{subsec:3d_method}), 2D projected separation and distance uncertainty sampling (Method 2, \ref{subsec:2d+3d_method}), and local density distribution and distance uncertainty sampling (Method 3, Sec. \ref{subsec:2dstat_3d_method}). Method 3, our novel approach, involves the use of the local density distribution of a star-forming region, under the assumption of isotropy, to derive the separation between stars in 3D space, given their 2D separation on the sky plane. The 3D separation is used to compute the FUV flux and its uncertainty.
The procedure and main results obtained in this study can be summarised as follow.
\begin{list}{$\bullet$}{}
    \item We selected a large sample of OBA-type stars, cross-referencing various stellar catalogues to include the contribution of late-type B and early-type A stars to the total FUV radiation, as described in Sec. \ref{sec:fuv_calculation:sample_star}.
    \\
    \item Using synthetic clusters, we verified that Method 1 can significantly underestimate FUV flux, while Method 2 leads to upper limit FUV fluxes (Sec. \ref{sec:synthetic_clusters}).
    Method 3, which consists in analytically prescribing the local density profile of star-forming regions, and sampling the derived PDF to obtain 3D separations given 2D separations (Sec. \ref{subsec:3d_method}), yields to a significantly larger fraction of FUV fluxes consistent within 1$\sigma$ error bars with true values, compared to the other methods.
    \\
    \item We provided a catalogue of the FUV flux (and uncertainty) at star-hosting discs in Taurus, $\rho$ Oph, UppSco, CrA, Lupus, ChamI, ChamII, ONC, $\sigma$ Ori, $\lambda$ Ori, NGC 1977, 25 Ori, NGC2024, and the sub-clusters of Serpens. This catalogue is available at the CDS. 
    \\
    \item 
    In regions where no O-type and early-type B stars are found, the irradiation induced by less massive stars (late-type B and early-type A) can even reach $\sim10^{2} \ \mathrm{G}_{0}$ (Sec. \ref{sec:results}). These FUV radiation levels are not negligible for disc evolution and planet formation, and therefore the impact of the environment must be taken into account when comparing discs in different regions.  
    \\
    \item We investigated the relation between FUV flux and dust disc mass, for discs in our sample presenting dust disc mass measurements in literature, considering the average age of their star-forming regions (Sec. \ref{subsec:dust_masses}). We underline the need for stellar mass and dust disc mass measurements at intermediate-to-high irradiation levels (>$10^{2} \ \mathrm{G}_{0}$), particularly for regions $\sim$1-3 Myr old (age range where most of the disc population is detected), to make reliable claims on a general correlation across different regions. 
    \\
    \item We included a study of the FUV attenuation caused by interstellar dust extinction (Appendix \ref{appendix:extinction}). We assumed that the FUV radiation emitted by the massive stars members of a certain region is attenuated by an average FUV extinction magnitude retrieved from dust extinction maps, while using an average galactic extinction value for the radiation from field massive stars. The resulting decrease in median FUV flux per region is not significant (a few $\mathrm{G}_{0}$, as shown in Table \ref{tab:extinction}). However, interstellar dust inside individual regions might be not resolved by dust maps, and therefore produce higher extinction than what assumed.  
\end{list}

This study shows how information provided by the 2D geometry of a star-forming region can be used to constrain 3D separations between star pairs within that region. We applied these estimates to compute the FUV flux at the position of nearby star-hosting discs. This approach can be extended to further regions in future studies.

\section{Data availability}
Tables \ref{tab:big_table_of_fluxes} in its extended version is only available in electronic form at the CDS 
or via \href{http://cdsweb.u-strasbg.fr/cgi-bin/qcat?J/A+A/}{http://cdsweb.u-strasbg.fr/cgi-bin/qcat?J/A+A/}.

\begin{acknowledgements}
RA and GR acknowledge funding from the Fondazione Cariplo, grant no. 2022-1217, and the European Research Council (ERC) under the European Union’s Horizon Europe Research \& Innovation Programme under grant agreement no. 101039651 (DiscEvol). Views and opinions expressed are however those of the author(s) only, and do not necessarily reflect those of the European Union or the European Research Council Executive Agency. Neither the European Union nor the granting authority can be held responsible for them.
AJW has received funding from the European Union’s Horizon 2020 research and innovation programme under the Marie Skłodowska-Curie grant agreement No 101104656. 
\end{acknowledgements}

\bibliographystyle{bibtex/aa} 
\bibliography{bibtex/mybib} 

\clearpage

\begin{appendix}
\section{Gallery of Investigated Star-forming Regions and Adopted Density Profiles}\label{Appendix:density_profiles}
\begin{figure*}[htbp]
    \centering
    \includegraphics[width=\textwidth]{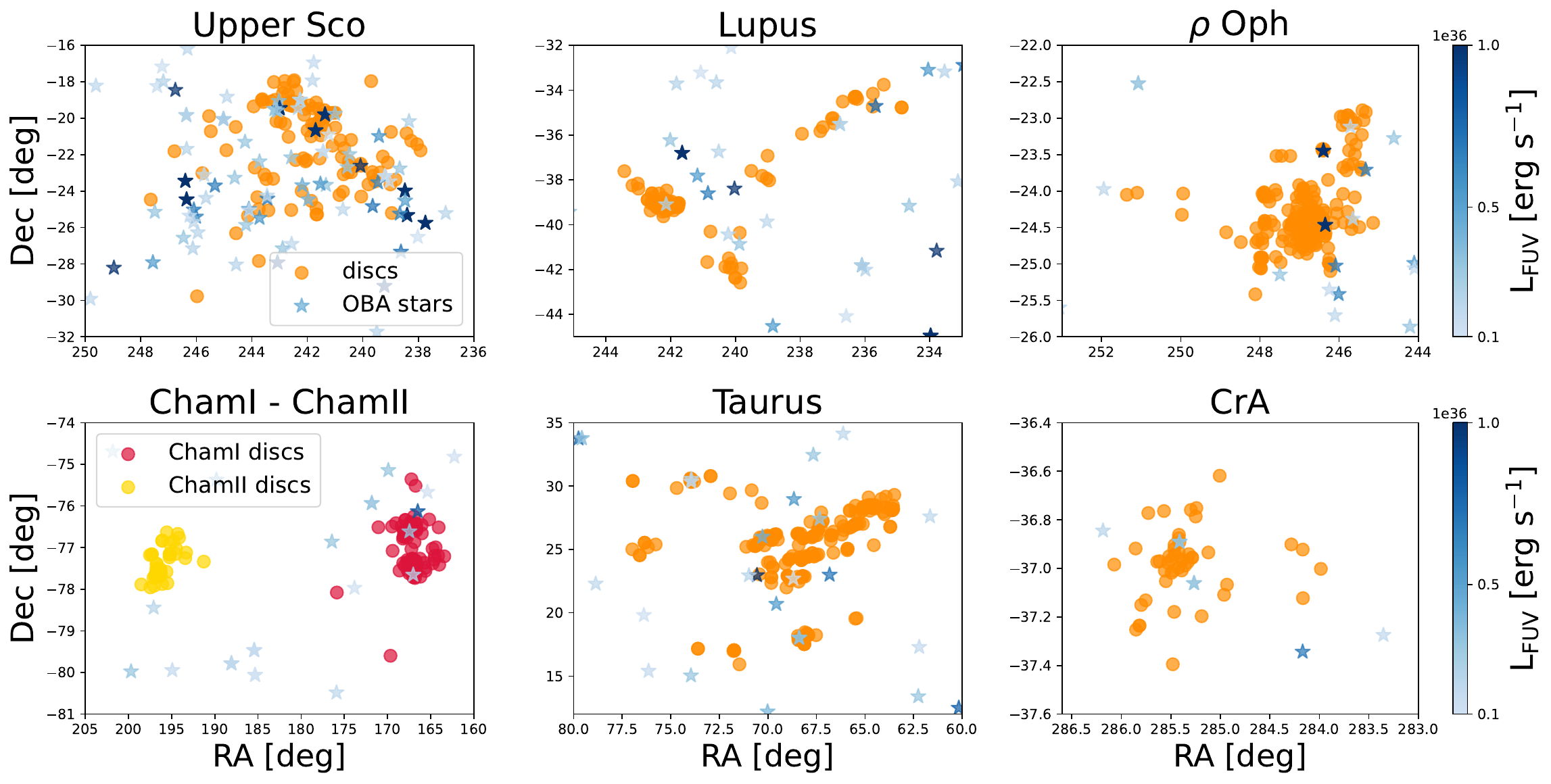}
    \caption{2D sky maps showing the positions (RA, Dec) of discs and neighbouring massive stars in the nearby star-forming regions included in this work. ChamI and ChamII are shown together due to their small relative distance.}
    \label{fig:2d_maps_nearby}
\end{figure*}
\begin{figure*}[htbp]
    \centering
    \includegraphics[width=\textwidth]{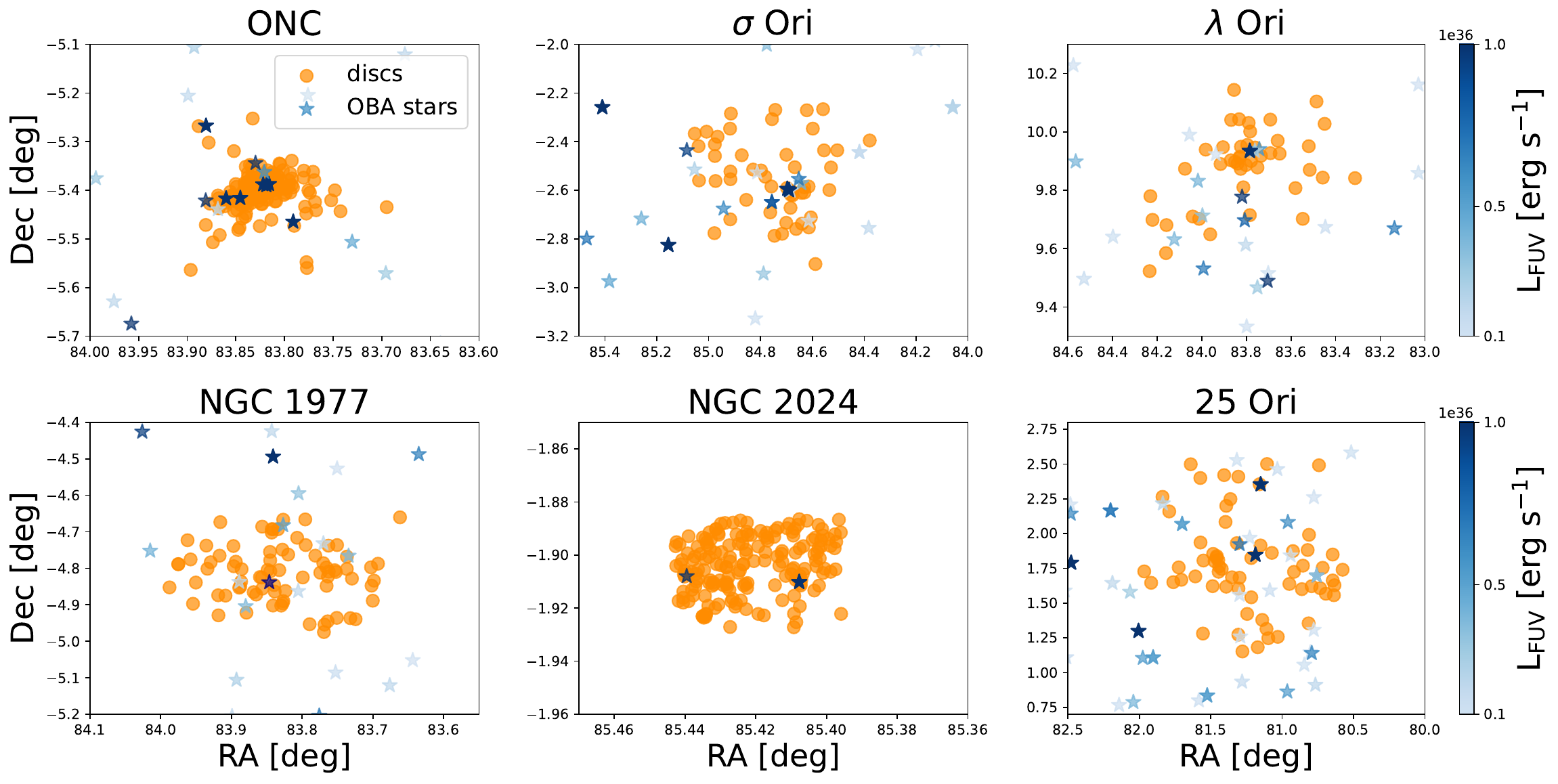}
    \caption{2D sky maps showing the RA and Dec position of the disc sample and neighbouring massive stars in the Orion regions.}
    \label{fig:2d_maps_orion}
\end{figure*}
\begin{figure*}[htbp]
    \centering
    \includegraphics[width=\textwidth]{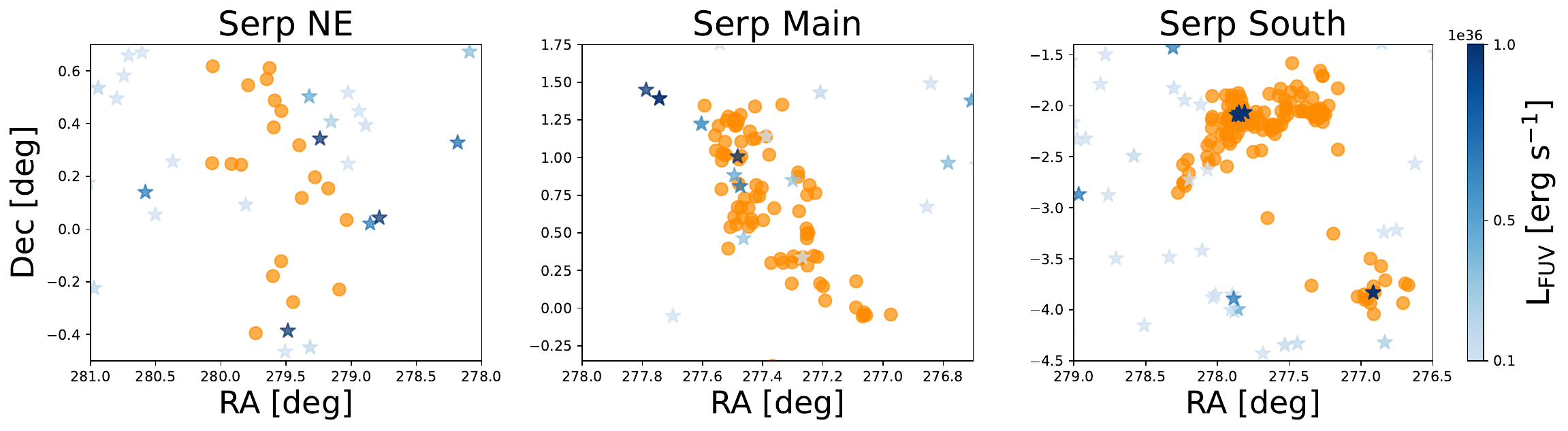}
    \caption{2D sky maps presenting the location (RA, Dec) of discs and neighbouring massive stars in the Serpens sub-regions.}
    \label{fig:2d_maps_serp}
\end{figure*}
\begin{table*}[h]
    \caption{\label{tab:density_profiles} Parameters of the volume density profiles used for the star-forming regions included in this work.}
    \centering
    \vspace{0.2cm}

    \begin{tabular}[H]{|P{2.0cm}| P{1.4cm}| P{1.4cm}|P{1.4cm}|P{1.4cm}|P{1.4cm}|P{1cm}|P{1.4cm}|}

    \hline
    \hline

    Region & $\alpha_{\mathrm{1}}$ & $\beta_{\mathrm{1}}$ & $r_{\mathrm{lim}}$ & $\alpha_{\mathrm{2}}$ & $\beta_{\mathrm{2}}$ & $r_{\mathrm{b}}$ & $\gamma$ \\

    \hline
    
    UppSco & 7.36E-05& 1.83& 1.5E-1& 2.16E-04& 6.04E-01& 12& 1.6\\
    Lupus &  1E-6& 2.6& 3E-2& 2E-5& 1.7& 18& 1.7\\
    $\rho$ Oph & 3E-5& 2.4& 1E-2& 1E-2& 0.8& 3& 1.2  \\
    Taurus & 2E-4& 2.2& 3E-2& 1.5E-3& 1.5& 20& 2.1 \\
    ChamI & 1.5E-4& 2.3& 2E-2& 6E-3& 1.3& 3.8& 1.5 \\
    CrA & 1E-4& 2.2& 5E-2& 2E-3& 1.22& 10& 1.8 \\
    ONC & - & - & - & 3E-2& 1.8& 3& 2.5 \\
    $\sigma$ Ori & - & - & - & 1E-2& 2.2& 8.0& 1.8 \\
    $\lambda$ Ori & - & - & - & 1E-2 & 2.3 & 10 & 2 \\
    NGC 1977 & - & - & 5E-3& 8E-3& 1.0& 4.0& 1.8 \\
    25 Ori & - & - & - & 3E-2& 0.9& 6& 1.5 \\
    Serp Main & - & - & 5E-2& 3E-3& 1.1& 8& 1.9 \\
    Serp NE & - & - & 6E-2& 2E-3& 1.05& 8& 2 \\
    Serp South & - & - & 1E-2& 1E-3& 1.2& 10& 1.3\\
    \hline
    
    \end{tabular}
    \tablefoot{ Parameters adopted for the volume density profiles of the regions investigated in this work. ChamII is not included as no relevant massive OBA type star is found near the studied stars-hosting discs. While, NGC2024 presents a sample of confirmed members which is too small and spread to reliably define a density profile. The listed parameters refer to Eq. \eqref{eq:rho_double_powerlaw} for UppSco, Lupus, $\rho$ Oph, Taurus, ChamI, and CrA. For the ONC, $\sigma$ Ori, and $\lambda$ Ori we used Eq. \eqref{eq:rho_single_powerlaw}, while NGC1977 and Serp regions are described by Eq. \eqref{eq:rho_double_powerlaw} at distances larger than $r_{\mathrm{lim}}$.}  
\end{table*}
The positions on the star-hosting discs considered in each star-forming region included in this work, and the location of their closest massive stars, are shown in Fig. \ref{fig:2d_maps_nearby}, \ref{fig:2d_maps_orion}, and \ref{fig:2d_maps_orion}, divided in regions in $\lesssim 200$ pc, Orion regions, and Serpens regions. The OBA stars are colour-coded based on their FUV luminosity, where more massive stars are associated with higher FUV luminosity.

In this work, we employed the Abel's theorem to derive the volume local density profile of star-forming regions, given an analytical prescription for the surface local density profile (method 3 in Sec. \ref{subsec:2dstat_3d_method}). As demonstrated in Sec. \ref{sec:synthetic_clusters}, the local density profile can be analytically described by a double power law with exponential cut off for substructured regions, or by a truncated single power law (or a Plummer profile) for centrally concentrated regions. The parameters for the power law profiles, as applied to the star-forming regions in this study, are listed in Table \ref{tab:density_profiles}. The terms reported in the Table refers to the expressions:
\begin{equation}
    \hat{\rho}(r) =
    \Bigg\{
    \begin{array}{ll}
        \alpha_{1} \ r^{-\beta_{1}}, & \text{if} \ r \leq r_{\mathrm{lim}}, \\
        \alpha_{2} \ r^{-\beta_{2}} \ \exp{ [ -(r/r_{\mathrm{b}})^{-\gamma} ] }, & \text{if} \ r > r_{\mathrm{lim}},
        \label{eq:rho_double_powerlaw}
    \end{array}
\end{equation}
where the 3D separation $r$ is normalized to 1 pc, $r_{\mathrm{lim}}$ is the limit distance around which the steepness of the power law profile changes, and $r_{\mathrm{b}}$ is the boundary radius setting the length scale of the star-forming region. This is valid for a region with internal substructures.
For the ONC, $\sigma$ Ori, $\lambda$ Ori, and 25 Ori regions, we employed a centrally concentrated geometry, with density profile that is function of the distance from the centre of the region, written as:
\begin{equation}
    \hat{\rho}(r) = \alpha_{2} \ r^{-\beta_{2}} \ \exp{ [ -(r/r_{\mathrm{b}})^{-\gamma} ] },
    \label{eq:rho_single_powerlaw}
\end{equation}
where $r$ is normalized to 1 pc and in Table \ref{tab:density_profiles} the terms $\alpha_{1}$, $\beta_{1}$, and $r_{\mathrm{lim}}$ assume null values as they do not enter in this expression. 

The accuracy in estimating the volume density profiles requires having a substantial number of stars per unit separation (i.e. a nearly complete stellar census of the star-forming regions), to properly account for the 2D internal geometry. 
In this study, we retrieved membership catalogues for each region from those available in literature.
Assigning the stellar membership requires accurate measurements of positions and proper motions, which becomes increasingly difficult in more distant regions, such as Orion and Serpens in our case. Therefore, the defined density profiles may need refinement as future, updated stellar censuses become available. 
The confirmed stellar membership of NGC 2024 (\citeauthor{APOGEE_survey} \citeyear{APOGEE_survey}, \citeauthor{SigmaOri_3D_2024} \citeyear{SigmaOri_3D_2024}) consists of stars too spatially separated with respect to the compact sample of discs considered in this work, to reliably make claims about the density profile per unit separation. Thus, our new proposed approach to evaluate the FUV flux cannot be applied in this region, which is not included in Table \ref{tab:density_profiles}. However, we provided FUV flux estimates using method 2 (Sec. \ref{subsec:2d+3d_method}), being aware that this is an overestimate of the true value.
The density profiles of NGC 1977 and Serpens sub-regions are described by single power law relations (where $\hat{\rho}$ is function of the separation between pairs on stars) for distances larger than a defined limit separation, $r_{\mathrm{lim}}$, due to the lack of a complete stellar membership at smaller separations, where the shape of $\hat{\rho}$ is primarily influenced by binary systems.
However, we verified that no pairs of star-hosting disc - massive star are present at separations smaller than this limit value, and therefore, the defined profile remains valid for our calculations.
The Cham II region does not contain any significant OBA stars, and therefore, we did not use the local density profile to evaluate the FUV flux in this region.

\section{Catalogue of the FUV Flux on Star-Bearing Disc in Nearby Regions}\label{appendix:big_table}
\begin{table*}[h]
\caption{\label{tab:big_table_of_fluxes} FUV flux at the position of the considered discs in 14 nearby star-forming regions.}
    \centering
    \vspace{0.2cm}

    \begin{tabular}[H]{|P{3.3cm} | P{1.6cm} | P{1.6cm} | P{1.6cm} | P{2.3cm} |P{2.3cm} |P{2.3cm} |}

    \hline
    \hline

    Disc name & Region & RA & Dec & FUV & FUV 16\% & FUV 84\% \\
    & & [deg] & [deg] & [$\mathrm{G}_{0}$] & [$\mathrm{G}_{0}$] & [$\mathrm{G}_{0}$] \\

    \hline
    
    J16083455-2211559 & UppSco & 242.1439 & -22.1988 & 19.0867 & 14.0405 & 24.2912 \\
    ... & ... & ... & ... & ... & ... & ... \\
    J16222521-2405139 & rho\_Oph & 245.6050 & -24.0872 & 13.7007 & 11.4185 & 15.9090 \\
    ... & ... & ... & ... & ... & ... & ... \\
    J05352536+1008383 & lambda\_Ori & 83.8556 & 10.1439 & 214.6073 & 81.4543 & 305.3879 \\
    ... & ... & ... & ... & ... & ... & ... \\

    \hline    
    \end{tabular}
    \tablefoot{ FUV flux best estimate and uncertainties shown as 16$^{\mathrm{th}}$ and 84$^{\mathrm{th}}$ percentiles of the distribution, at the position of star-hosting discs in the 14 regions included in this work. The complete is available at the CDS. } 
    
\end{table*}
The approach presented in this study allows to accurately estimate FUV flux (and its uncertainty). We applied this method to evaluate the FUV flux at the position of star-bearing discs in 14 nearby star-forming regions, compiling a catalogue of the FUV flux for a large sample of stellar objects. The complete table containing the flux best estimates is available at the CDS.
Few rows of the table, with specific content of the columns, are shown in Table \ref{tab:big_table_of_fluxes}. For each star-hosting disc, we included the corresponding star-forming region, RA and Dec positions from Gaia DR3, median best estimate FUV flux, and uncertainties as 16$^{\mathrm{th}}$ and 84$^{\mathrm{th}}$ percentiles of the distribution.

\section{The Role of the Extinction}\label{appendix:extinction}
\begin{table*}[h]
    \caption{\label{tab:extinction} 
    Average interstellar extinction used for each region, and percentage of decrease in FUV flux due to attenuation of radiation from stars members of the region and field stars.}
    \centering
    \vspace{0.3cm}

    \begin{tabular}[H]{|P{2.3cm}| P{2cm}| P{2.7cm}|P{2.7cm}|P{2.3cm}|P{2.3cm}|}

    \hline
    \hline

    Region & $A_{\mathrm{FUV}}$ & median $F_{\mathrm{FUV, \ comp}}$ & median $F_{\mathrm{FUV, \ corr}}$ & Reduction & Reduction  \\   
     &  [mag kpc$^{-1}$] & [$\mathrm{G}_{0}$] & [$\mathrm{G}_{0}$] & internal FUV $\%$ & field FUV $\%$ \\

    \hline

    UppSco & 5.31 & 20.11 & 17.69 & 3.0 & 24.6 \\
    Lupus & 6.15 & 3.48 & 2.15 & 13.3 & 43.4 \\
    $\rho$ Oph & 11.59 & 18.33 & 16.81 & 1.86 & 19.7 \\
    Taurus & 0.56 & 2.92 & 1.54 & 0.5 & 48.5 \\
    ChamI & 1.21 & 4.57 & 3.19 & 0.19 & 45.3 \\
    ChamII & 1.12 & 2.76 & 1.37 & 0.0 & 50.5\\
    CrA & 1.47 & 9.94 & 8.36 & 0.12 & 43.3 \\
    ONC & 1.23 & 33713.9 & 33380.1  & 0.022 & 25.02 \\
    $\sigma$ Ori & 1.28 & 232.5 & 227.9 & 0.22 & 13.6 \\
    $\lambda$ Ori & 1.03 & 536.9 & 531.5 & 0.12 & 41.6 \\
    NGC 1977 & 1.24 & 100.1 & 98.2 & 0.12 & 10.1 \\
    25 Ori &  0.83 & 18.14 & 14.75 & 0.42 & 44.7\\
    Serp Main & 0.56 & 3.93 & 2.66 & 0.13 & 70.6\\
    Serp NE &  0.89 & 2.79 & 1.51 & 0.37 & 75.2 \\
    Serp South & 1.12 & 44.27 & 42.98 & 0.40 & 72.2 \\
    \hline
    
    \end{tabular}
    \tablefoot{ For each region in the first column, the average interstellar dust extinction magnitude converted to the FUV band is presented in the second column. Column 3 and 4 contain the median FUV flux as computed in this work and corrected by the extinction. Column 5 and 6 distinguish between the decrease in FUV flux due to dust extinction between the internal and external OBA stars of single regions. An average value of $\sim$2.1 mag kpc$^{-1}$ is applied to the incident FUV flux from field stars. }  
\end{table*}
Interstellar gas and, more importantly, dust can efficiently shield protoplanetary discs from an incoming UV radiation. 
Quantify the relevance of this process is challenging, as it requires knowledge of the internal geometry of a star-forming region and the column density between discs and massive stars.
Moreover, gas outflows produced by massive stars can reduce the attenuating material, contributing to the variation in the irradiation experienced by protoplanetary discs with time (e.g. \citeauthor{Bania_Lyon_1980} \citeyear{Bania_Lyon_1980}, \citeauthor{Jeffreson_2021} \citeyear{Jeffreson_2021}).
Simulations of molecular clouds show that discs in high-mass clouds containing several O-type stars, can be efficiently shielded by FUV radiation during the first $\sim$0.5 Myr of their evolution (\citeauthor{Ali_2019} \citeyear{Ali_2019}, \citeauthor{Qiao_extinction_2022} \citeyear{Qiao_extinction_2022}). After this shielding time, discs are rapidly subject to higher FUV fluxes.
In order to investigate the role of the interstellar dust in attenuating the FUV irradiation, we used the 3D dust extinction map provided by \citet{Dust_extinction_map_22}, which uses Gaia and 2MASS photometric data to covers a volume of $\sim$ 6 kpc x 6 kpc x 800 pc around the Sun, reaching resolution of few pc. As the resolution limit of this map is too large to resolve cloud substructures and accurately quantify the column density of material between star-hosting discs and massive stars in a same star-forming region, where the separation between them is smaller than $\sim$5 pc, we retrieve from the 3D dust map an average extinction magnitude value for each region. This value is included in the calculation of the FUV flux produced by the massive stars internal to a certain region to the star-hosting discs of the same region. We converted visual magnitudes, provided by the dust map, in FUV magnitudes using a factor $A_{\mathrm{FUV}}/A_{\mathrm{V}} \simeq 2.7$ (\citeauthor{Cardelli_1989} \citeyear{Cardelli_1989}, \citeauthor{Extinction_review} \citeyear{Extinction_review}).
For the field massive stars (those outside the considered region) we assumed an average galactic extinction value of $\sim 3 A_{\mathrm{V}} \simeq 2.1$ mag kpc$^{-1}$, as in the work of \citet{cleeves_imlup}. 
The individual FUV flux corrected for the UV extinction magnitude is:
\begin{equation}
    F_{\mathrm{FUV, \ corr}} = F_{\mathrm{FUV, \ comp}} \ 10^{-\frac{A_{\mathrm{FUV}}}{2.5}},
    \label{eq:F_corr_ext}
\end{equation}
where $F_{\mathrm{FUV, \ comp}}$ is our flux best estimate and $A_{\mathrm{FUV}}$ is the average extinction magnitude in the FUV range of wavelengths.
In Table \ref{tab:extinction} we show the average extinction magnitude for each region, derived from the 3D dust extinction map and converted to FUV wavelengths, the median FUV flux computed using the local density distribution (Method 3, Sec. \ref{subsec:2dstat_3d_method}), which provides our best estimate of the flux, and the flux value corrected by the UV extinction. We distinguish between the percentage of flux reduction caused by massive hot stars internal the star-forming region (members) and external (field stars). In general, the median FUV flux per region is not drastically reduced by extinction. Since massive stars internal to a region are located closer to discs than field massive stars, their irradiation is less attenuated.
In regions like the ONC, where the FUV flux is dominated by few massive stars very close to discs, the extinction correction to the total FUV flux is minimal. In these cases, even though the FUV flux reduction from field stars can be significant ($\sim$25 \% for the ONC, $\sim$40 \% for $\lambda$ Ori), the total FUV flux decrease is negligible ($\lesssim$1\%).

We provided a tentative estimate of the impact of the dust extinction in attenuating FUV photons incident on disc-hosting stars assuming, for the disc-hosting stars of a certain region, that the FUV radiation emitted by massive stars members of that region is attenuated by the average UV extinction magnitude provided by dust extinction maps at that location, and applying an average galactic extinction to the FUV radiation emitted by massive stars outside the region. The results presented in Table \ref{tab:extinction} show that the median FUV flux per region is not significantly reduced by extinction. 
However, we highlight that the extinction magnitudes used are average values based on a dust map with low resolution for our scopes, meaning that intra-star extinction due to dust material within a single region might be higher than what assumed. Future improvement will be possible with the availability of 3D high-resolution dust extinction maps of star-forming regions. 

\section{Tentative Relation Between Dust Disc Mass and FUV Flux Across Regions}\label{appendix:corr_mdust_fuv}
\begin{figure}[htbp]
    \centering
    \includegraphics[width=0.45\textwidth]{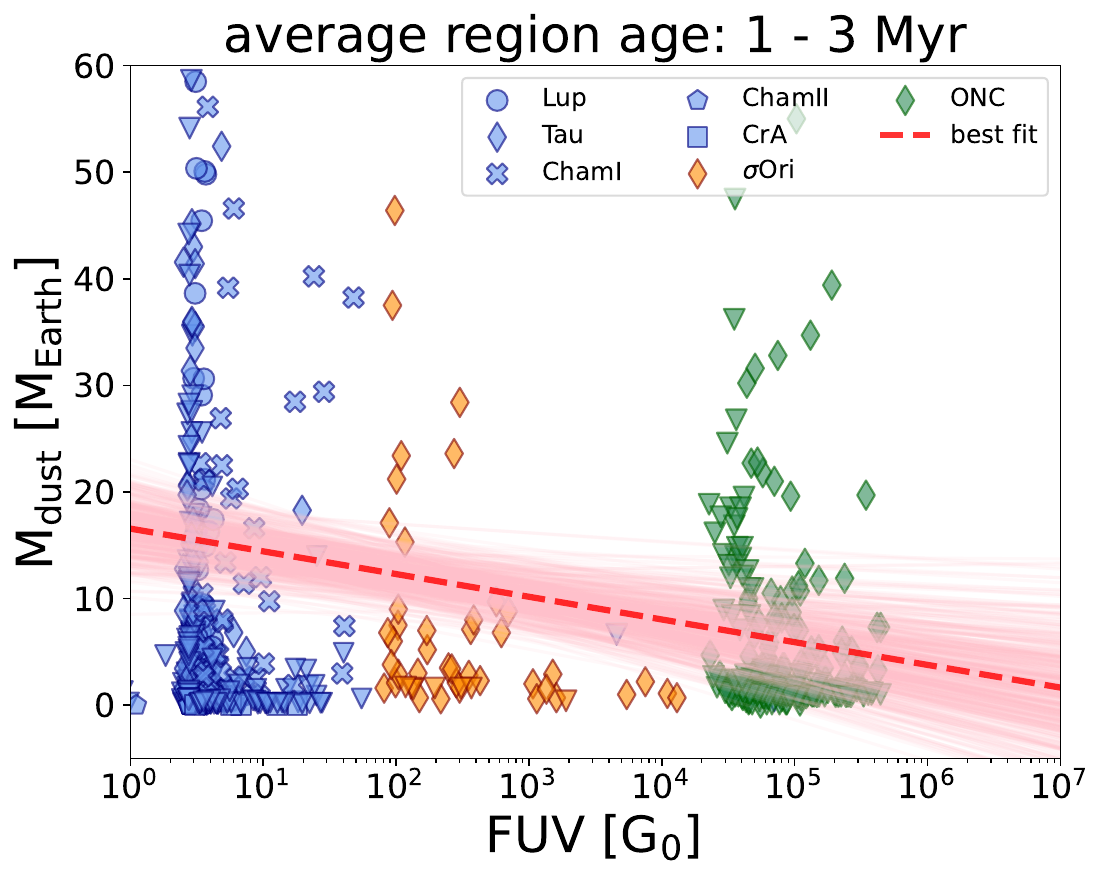}
    \hspace{0.04\textwidth}
    
    \caption{Tentative correlation between disc dust masses, and FUV flux best estimate in regions with average age 1-3 Myr. Triangles indicate upper limit dust disc masses. Uncertainties in masses and flux, and upper limits, are included in the derivation of the best fit model and not displayed for more clarity in the Figure. }
    \label{fig:corr_fuv_mdust_mstar}
\end{figure}
A clear dependence of the disc dust mass (i.e. observed mm flux emission) on the FUV radiation has been observed across individual star-forming regions, 
such as the ONC \citep{Eisner_2018_onc}, $\sigma$ Ori \citep{Mauco_2023_sori}, Serpens (van Terwisga et al. in prep.), and the extended population of Orion targeted by the SODA Survey \citep{SODA_2023}. For what concerns the SODA disc sample, where a correlation is claimed in \citet{SODA_2023}, we are unable to provide a more FUV flux estimate because, in the extended population of Orion, a confirmed sub-cluster division (essential to apply the method proposed in this work), is not currently defined. Future sub-cluster division and membership census of Orion will make possible a better FUV flux estimate also for the disc sample in the extended population.
We explored the dependence of the dust disc mass from the FUV flux across various regions using the updated FUV flux values and uncertainties. We focus on regions with average age between $\sim$1 and $\sim$3 Myr, which is the age range in which statistics is expected to be more robust as more than two regions of our sample present dust disc mass measurements.
The result is shown in Fig. \ref{fig:corr_fuv_mdust_mstar}.
Accounting for upper limits in dust musses and including uncertainties on both mass and FUV flux, we found a negative correlation:
\begin{equation}
    \frac{M_{\mathrm{dust}}}{M_{\mathrm{Earth}}} = (16.25 \pm 2.26 ) + \log_{10}\left( \frac{F_{\mathrm{FUV}}}{1 \mathrm{G}_{0}} \right) \ (-2.04 \pm 0.79) ,
    \label{eq:corr_mdust_fuv}
\end{equation}
We highlight that the found correlation, described by Eq. \eqref{eq:corr_mdust_fuv}, is only tentative. Indeed, as discussed in Sec. \ref{subsec:dust_masses}, the number of dust mass measurements in the flux range $10^{2} - 10^{4} \ \mathrm{G}_{0}$, for discs in regions with average age 1-3 Myr, is not statistically comparable with the observations in the most nearby regions. Moreover, stellar masses are not measured for the majority of the intermediately and highly irradiated objects included in this study, with a bias toward massive stars among the measurements. Since the dependence of dust disc mass on stellar mass (e.g. \citeauthor{Manara_2023} \citeyear{Manara_2023}) should be considered in this analysis, we refer to future work for a more accurate investigation.

\end{appendix}

\end{document}